\documentclass[letterpaper,12pt]{extarticle}
\usepackage{showkeys}
\usepackage{graphicx}
\usepackage{dcolumn}
\usepackage{bm}
\usepackage{tabularx} 
\usepackage{preamble}
\usepackage{circledsteps}
\usepackage[utf8]{inputenc}
\usepackage{overpic}
\usepackage{circuitikz}
\usepackage[export]{adjustbox}
\usepackage{upgreek}
\usepackage{enumerate}
\usepackage{url}
\usepackage{bbm}
\usepackage{setspace}
\usepackage[font=small,labelfont=bf]{caption}
\usepackage{subfig}
\usetikzlibrary{calc}
\usepackage{extarrows}
\usepackage{amsmath,amssymb}
\usepackage{booktabs,amsfonts,dcolumn}
\usepackage{circledsteps}
\usepackage{subcaption}
\usepackage{booktabs,subcaption,amsfonts,dcolumn}
\usetikzlibrary{decorations.pathmorphing}
\newcommand{\DD}{\mathsf{D}}
\newcommand{\EE}{\mathsf{E}}
\numberwithin{equation}{section}
\newcolumntype{d}[1]{D..{#1}}

\definecolor{refkey}{rgb}{0.9451,0.2706,0.4941}
\definecolor{labelkey}{rgb}{0.9451,0.2706,0.4941}
\usepackage{cite}
\usepackage{simplewick}

\newcommand{\fixed}{\texttt{fixed}}
\def\z2{$\mathbb{Z}_2$}

\definecolor{darkgray}{rgb}{0.33, 0.33, 0.33}
\usepackage{braket}
\newcommand{\kket}[1]{\left|\left.#1\right\rangle\!\right\rangle}

\newcommand{\cket}[1]{|#1)}
\newcommand{\cbra}[1]{(#1|}
\newcommand{\cbraket}[1]{(#1)}
\newcommand{\Tr}{{\mathrm{Tr}}}

\newcommand{\qbinom}[2]{\genfrac{[}{]}{0pt}{}{#1}{#2}_q}
\newcommand{\punch}[1]{\textcolor{NavyBlue}{#1}}
\usepackage[left=30mm,right=30mm,top=30mm,bottom=30mm]{geometry}
\usetikzlibrary{arrows.meta,calc}
\usepackage{authblk}
\numberwithin{equation}{section}
\definecolor{MyGreen}{RGB}{10,120,10}
\usetikzlibrary{shapes.misc}
\tikzset{cross/.style={cross out, draw=black, minimum size=2*(#1-\pgflinewidth), inner sep=0pt, outer sep=0pt},
cross/.default={1pt}}
\usepackage{diagbox}
\usepackage{siunitx}
\usetikzlibrary{positioning}
\usepackage{cjhebrew}

\DeclareFontFamily{U}{rcjhbltx}{}
\DeclareFontShape{U}{rcjhbltx}{m}{n}{<->rcjhbltx}{}
\DeclareSymbolFont{hebrewletters}{U}{rcjhbltx}{m}{n}

\let\aleph\relax\let\beth\relax
\let\gimel\relax\let\daleth\relax

\DeclareMathSymbol{\aleph}{\mathord}{hebrewletters}{39}
\DeclareMathSymbol{\beth}{\mathord}{hebrewletters}{98}
\DeclareMathSymbol{\gimel}{\mathord}{hebrewletters}{103}
\DeclareMathSymbol{\daleth}{\mathord}{hebrewletters}{100}

\DeclareMathSymbol{\lamed}{\mathord}{hebrewletters}{108}
\DeclareMathSymbol{\mem}{\mathord}{hebrewletters}{109}
\DeclareMathSymbol{\ayin}{\mathord}{hebrewletters}{96}
\DeclareMathSymbol{\tsadi}{\mathord}{hebrewletters}{118}
\DeclareMathSymbol{\qof}{\mathord}{hebrewletters}{113}
\DeclareMathSymbol{\shin}{\mathord}{hebrewletters}{152}
\definecolor{DarkEverGreen}{RGB}{70,150,70}

\newcommand{\revisionaddition}[1]{{\color{red}#1}}

\begin{document}
\onehalfspacing
\title{\Huge
A JT/KPZ correspondence
}

\author{Masataka Watanabe\footnote{\href{mailto:max.washton@gmail.com}{max.washton@gmail.com}}
}
\affil{\it \small The University of Tokyo, Tokyo 113-0033, Japan}
\date{}
\pagenumbering{gobble}

\maketitle

\thispagestyle{empty}
\begin{abstract}
\noindent
    We point out a correspondence between the Jackiw--Teitelboim (JT) gravity and the stationary measure of the Kardar--Parisi--Zhang (KPZ) equation on an interval.
    By relating the Schwarzian limit of the double-scaled SYK to the weakly asymmetric limit of the open ASEP, we establish that the path-integral measure defining the Euclidean evolution between two end-of-the-world branes in JT gravity can be interpreted as the stationary measure of the KPZ equation on an interval with Neumann boundary conditions.
    We also establish the match between correlation functions.

\end{abstract}
\newpage
\pagenumbering{arabic}
\pagestyle{plain}
\onehalfspacing
\tableofcontents
\newpage

\setcounter{tocdepth}{2}

\section{Introduction}

Imagine putting flame to a paper sheet;
watch the fire spread with time.
As physicists, we might be tempted to describe the propagation of the fire front with an evolution equation. 
Surprisingly, such a simple everyday phenomenon turns out to be a remarkably rich subject.

Experiments have shown \cite{maunuksela1997kinetic}
that fire-front propagation can be effectively modelled by a stochastic differential equation known as the Kardar--Parisi--Zhang (KPZ) equation \cite{kardar1986dynamic}.
It describes an evolution of the height profile of the fire-front $h(t,x)$, a function of time $t$ and the spatial coordinate $x$, as
\begin{align}
\partial_t h
= \nu {\partial_x^2}h +\frac{\lambda}{2} \left(\partial_x h\right)^2 + \sqrt{D}\xi(x,t),
\end{align}
where $\xi$ is a spacetime white noise satisfying $\braket{\xi(t,x)\xi(t^\prime,x^\prime)}=\delta(t-t^\prime)\delta(x-x^\prime)$.
By rescaling units, we will set $\nu=\lambda=D=1$ hereafter.
The equation is universal in the sense that it describes a broad class of driven interfaces, such as bacterial-colony expansion \cite{wakita1997self,hallatschek2007genetic}.

Given a stochastic equation, one is typically interested in its late-time behaviour. 
For example, for the KPZ equation on a full line, the stationary measure was shown to be given by a two-sided Brownian motion \cite{bertini1997stochastic,funaki2015kpz}. 
Furthermore, the probability distribution of the height can also be computed at late times \cite{sasamoto2010one,amir2011probability,calabrese2011exact,imamura2012exact,imamura2013stationary}.

In terms of experiments, perhaps a more realistic situation would be to describe surface evolutions on a bounded space with boundary conditions.
The open KPZ equation is a variant of the KPZ equation defined on an interval $0\leq x \leq X$ with Neumann boundary conditions.
We will henceforth be interested in the stationary measure for the open KPZ equation, which will turn out to be completely different from the counterpart defined on a full line.

Discretization is a typical way of tackling such a problem.
In \cite{corwin2018open,parekh2019kpz,yang2025kpz}, it was shown that the open KPZ can be obtained as a weakly asymmetric limit of the open asymmetric simple exclusion process (ASEP). 
ASEP is among the famous solvable models of driven diffusive systems, and has been extensively studied in various contexts including physics, social science, and biology \cite{derrida1998exactly,macdonald1968kinetics,schutz2001exactly,schadschneider2010stochastic}. 
It is a continuous-time Markov process of particles on a one-dimensional lattice (made of $N$ sites) that hop asymmetrically, equipped with boundary inflow and outflow (to be defined in the main body of the text).
Denoting the asymmetry parameter as $q$ (to be defined in the main body of the text), the claim of \cite{corwin2018open,parekh2019kpz,yang2025kpz} is that the height function of the open KPZ can be obtained as a suitable limit of particle configurations of ASEP, by taking $N\to\infty$ while scaling $q=e^{-2/\sqrt{N}}\to 1$ and boundary inflow/outflow rates suitably.

The problem of finding the stationary measure of the open KPZ equation is then reduced to find that of ASEP.
Fortunately, it can be derived algebraically because of the underlying integrability structure of ASEP -- it can be computed using the matrix product ansatz \cite{derrida1993exact}.
The procedure starts as \emph{assuming} that the stationary distribution of ASEP can be represented as a matrix product,
\begin{align}
    P(\vec{\tau})=\frac{\braket{W|\prod_{j=1}^{K}\left(\tau_j\DD + (1-\tau_j)\EE \right)|V}}{\braket{W|(\DD + \EE)^N|V}}.
\end{align}
This can then be checked \emph{a posteriori} to be indeed the case when
\begin{align}
    \DD \EE - q \EE \DD = \zeta(\DD + \EE),\quad
    \bra{W}(\alpha \EE - \gamma \DD)=\zeta \bra{W},\quad
    (\beta \DD - \delta \EE)\ket{V}=\zeta\ket{V}.
\end{align}
We will pick a convention where $\zeta\equiv \sqrt{1-q}$ throughout this paper.
The algebra, called the DEHP algebra, is then related to tri-diagonal matrices and $q$-orthogonal polynomials, whose power is responsible for the fact that various physical quantities can be computed in the ASEP stationary state \cite{uchiyama2004asymmetric,sasamoto1999one}.

Let us take a sudden turn here; \punch{\emph{all of the above has a parallel context in a theory of gravity in two dimensions, called the Jackiw-Teitelboim (JT) gravity}}.
JT gravity is a theory of two-dimensional gravity coupled to matter \cite{Jackiw:1984je,Teitelboim:1983ux}, and in the holographic context can be described by 
a $1d$ Schwarzian theory on the boundary of the manifold on which it lives \cite{Maldacena:2016upp}.
Once gauge-fixed, it is reduced to $1d$ Liouville quantum mechanics on coordinate $\phi$ with potential given by $V(\phi)\propto e^{-\phi}$ \cite{Bagrets:2016cdf}.
One can then think of $\phi$ as a renormalised length between the two boundaries in JT gravity \cite{Lam:2018pvp,Goel:2018ubv,Berkooz:2022fso,Lin:2022rbf}.

Meanwhile, the Sachdev-Ye-Kitaev (SYK) model is a disordered chaotic quantum mechanical model of $\mathtt{N}\gg 1$ Majorana fermions with all-to-all $\mathtt{p}$-body random interactions \cite{Sachdev:1992fk,Maldacena:2016hyu,Kitaev2015Talks}.
At low energies, the SYK model is also described by the Schwarzian theory and hence it is holographically dual to the JT gravity \cite{Maldacena:2016upp}.
The SYK model has a solvable double-scaling limit when $\lambda\equiv 2\mathtt{p}^2/\mathtt{N}$ is held fixed while taking $\mathtt{N}\to\infty$, known as the double-scaled SYK \cite{Berkooz:2018qkz,Berkooz:2018jqr,Berkooz:2024lgq}.
By taking $\mathtt{q}\equiv e^{-\lambda}\to 1$ while taking some continuous limit, it has been shown that the model reduces to 
$1d$ Liouville quantum mechanics, \emph{i.e.}, the ordinary SYK model and the JT gravity \cite{Berkooz:2018qkz,Berkooz:2018jqr,Lin:2022rbf,Berkooz:2024lgq}.

Our claim is that what the double-scaled SYK model is to the SYK model (or equivalently, the JT gravity) is what ASEP is to the open KPZ equation.
The double-scaled SYK can be solved by using a combinatorial technique known as the chord diagram technique, which then allows us to express physical quantities by using the transfer matrix $T=\DD+\EE-2/\sqrt{1-\mathtt{q}}$, with $\DD$ and $\EE$ satisfying the DEHP algebra as in ASEP \cite{Okuyama:2023byh,Watanabe:2024vad,Berkooz:2025ydg,Okuyama:2025ebn}.
By leveraging such hints, we first find that the path-integral measure defining the double-scaled SYK can be exactly matched with the stationary measure of ASEP.\footnote{More precisely speaking, the double-scaled SYK in question is a variant with $\mathcal{N}=2$ supersymmetry \cite{Berkooz:2020xne}. We will discuss such subtleties in the main body of the text.}
We then find that the same $\mathtt{q}\to 1$ limit which reduced ASEP to the open KPZ equation reduces the double-scaled SYK to the Liouville quantum mechanics.
This allows us to match the path-integral measure of JT gravity to the stationary measure of the open KPZ equation.

We will also match correlators between the two. 
Certain $2n$-point functions of the double-scaled SYK will be shown to exactly match $(n+1)$-point functions of ASEP stationary state.
Taking the same $\mathtt{q}\to 1$ limit, we will also see that thermal $2n$-point functions in JT gravity can be matched to stationary state $(n+1)$-point functions in open KPZ equation.

Hopefully, our JT/KPZ correspondence has interesting consequences, aside from connecting two completely different subjects.
It first of all constitutes that the stationary measure of the open KPZ equation can be obtained using Liouville quantum mechanics, which is known to describe the JT gravity.
The same observation was (unfortunately for us, fortunately for the world) already made in \cite{barraquand2021steady,corwin2024stationary,barraquand2023stationary}, but ours offers another route to finding Liouville quantum mechanics in the open KPZ.
We can also think about it as adding a corner to the rich world of $q$-orthogonal polynomials, one of which has also been unravelled recently to relate the double-scaled SYK to the Schur index of $4d$ $\mathcal{N}=2$ $SU(2)$ gauge theories \cite{Gaiotto:2024kze,Lewis:2025qjq,Berkooz:2025ydg}.\footnote{See also \cite{Gorsky:2018jdj,Gorsky:2020lma,Gorsky:2024xnl} for some other connections to KPZ universality and tri-diagonal matrices.}

The rest of the paper is organised as follows.
We start in Section \ref{sec:DEHP} by presenting the DEHP algebra and the $q$-deformed oscillator algebra, which will be the underlying algebraic structure which enables us to solve the ASEP and double-scaled SYK.
We then go on to study the ASEP and the weakly asymmetric limit to the open KPZ, focusing on their stationary measures in Section \ref{sec:asep}.
In Section \ref{sec:dssyk}, we study double-scaled SYK and the triple-scaling limit to JT gravity, after which we point out the correspondence between double-scaled SYK and the ASEP as well as between JT gravity and the KPZ equation in Section \ref{sec:correspondence}.
Finally, we conclude and discuss open questions in Section \ref{sec:outlook}.

\paragraph{Contributions}

Our main contribution in this work is the very matching of physical quantities between ASEP and the double-scaled SYK, as well as discovering that the weakly-asymmetric limit for the former and the triple-scaling limit for the latter are exactly the same limit due to the correspondence (Section \ref{sec:correspondence} and Appendices  \ref{sec:correlators}, \ref{sec:fin}).
Building up towards the correspondence, we review various aspects of ASEP and the double-scaled SYK.
In order to explain the correspondence smoothly, we sometimes depart from the usual way in which each topic is presented, or include new results.
Such sections are Section \ref{sec:twolayer} where we rewrite the known two-layer description of the ASEP stationary measure in terms of a new basis, as well as removing the Liggett's condition assumed in \cite{barraquand2023stationary}, and Section \ref{sec:triple} where the triple-scaling limit of ASEP is taken rather unconventionally.



\section{DEHP algebra}
\label{sec:DEHP}


\subsection{DEHP algebra}

We start with an underlying theme for the present paper, the DEHP algebra.
It was originally introduced when deriving the stationary measure of open ASEP \cite{Derrida:1992vu}.
It is made of two matrices ($\DD$ and $\EE$) and two vectors ($\ket{W}$ and $\bra{V}$), which satisfy 
\begin{align}
    \begin{gathered}
        \DD \EE - q \EE \DD = \sqrt{1-q}(\DD + \EE),\\\quad
        \bra{W}(\alpha \EE - \gamma \DD)=\sqrt{1-q} \bra{W},\quad
        (\beta \DD - \delta \EE)\ket{V}=\sqrt{1-q} \ket{V}.
    \end{gathered}
    \label{eq:DEHP}
\end{align}

It is immediate to see that it is related to the Arik--Coon $q$-oscillator algebra \cite{Arik:1973vg} \emph{via}
\begin{align}
    \DD \equiv \frac{1}{\sqrt{1-q}}+{a},\quad 
    \EE \equiv \frac{1}{\sqrt{1-q}}+{a^\dagger},
\end{align}
where $a^\dagger$ and $a$ satisfy the $q$-commutation relation,
\begin{align}
    aa^{\dagger}-qa^\dagger a=1.
\end{align}

Let us pick a natural basis for the DEHP algebra (or, equivalently, the $q$-oscillator algebra),
\begin{align}
    \begin{gathered}
        a^\dagger\ket{n}\equiv \sqrt{\frac{1-q^{n+1}}{1-q}}\ket{n+1}\\
        a\ket{n}\equiv \sqrt{\frac{1-q^n}{1-q}}\ket{n-1}, \quad a\ket{0}=0.
    \end{gathered}
    \label{eq:aaa}
    \end{align}
We also define an operator $\hat{N}$ satisfying $\hat{N}\ket{n}=n\ket{n}$.
We will see later that $\ket{n}$ corresponds to the chord state with $n$ chords in double-scaled SYK; we call the basis as the \emph{chord number basis}.
We also hereby define the $q\to 0$ limit of the $q$-deformed oscillator for later convenience:
\begin{align}
    \begin{gathered}
        b^\dagger \ket{\chi}\equiv \sqrt{\frac{1}{1-q}}\ket{\chi+1},\quad
        b\ket{\chi}\equiv \sqrt{\frac{1}{1-q}}\ket{\chi-1}.
    \end{gathered}
    \label{eq:bbb}
\end{align}

The operator $a^\dagger +a$ is symmetric and hence can be explicitly diagonalised \cite{Berkooz:2024lgq}.
Its spectrum is continuous and is parametrised by $0\leq \theta \leq \pi$, where we have
\begin{align}
        (a^\dagger+a)\ket{\theta}&=\frac{2\cos\theta}{\sqrt{1-q}}\ket{\theta}.\label{eq:qqqqqqq}\\
        \braket{n|\theta}&=\sqrt{\frac{(q,e^{\pm 2i\theta};q)_{\infty}}{2\pi(q;q)_n}}H_n(\cos \theta|q),
    \label{eq:coordinate}
\end{align}
where $(q;q)_n\equiv \prod_{i=1}^{n}(1-q^i)$, $(q,e^{\pm 2i\theta};q)_{n}\equiv (q;q)_n(e^{2i\theta};q)_n(e^{-2i\theta};q)_n$, and $H_n(\cos \theta|
q)$ is the continuous $q$-Hermite polynomial, all of which are introduced in Appendix \ref{sec:q-pol}.

\subsection{Coherent states}

In our basis, $\bra{V}$ and $\ket{W}$ can be expressed as a (generalised) $q$-coherent state.
Let us define
\begin{align}
    (\beta \DD - \delta \EE)\cket{\beta,\delta}=\sqrt{1-q}\cket{\beta,\delta}.
\end{align}
so that $\ket{W}=\cket{\beta,\delta}$ and $\bra{V}=\cbra{\alpha,\gamma}$.
Writing
\begin{align}
    \cket{\beta,\delta}=\sum_{n=0}^{\infty}c_n\ket{n},
\end{align}
we can see that $c_n$ satisfies a certain three-term recurrence relation, 
\begin{align}
    \left(1-\frac{\beta-\delta}{1-q}\right)c_n=\frac{\beta}{\sqrt{1-q}}\cdot\sqrt{\frac{1-q^{n+1}}{1-q}}c_{n+1}-\frac{\delta}{\sqrt{1-q}}\cdot\sqrt{\frac{1-q^{n}}{1-q}}c_{n-1}.
    \label{eq:3term}
\end{align}
From this, we can see that
\begin{align}
    c_n\propto \kappa_+(\beta,\delta)^n
    \label{eq:scaling}
\end{align}
at large $n$, where
\begin{align}
    \kappa_\pm(\beta,\delta)\equiv \frac{1-q-\beta+\delta\pm\sqrt{(1-q-\beta+\delta)^2+4\beta\delta}}{2\beta}.
    \label{eq:kappa}
\end{align}
Note that $\cket{\beta,\delta}$ is not normalisable unless $\kappa_+(\beta,\delta)<1$.
However, as we will see, for our purposes it is sufficient that they make sense in the correlator, so we just need $\braket{V|W}<\infty$.
This is ensured by setting $\kappa_{+}(\beta,\delta)\kappa_+(\alpha,\beta)<1$.

When $\delta =0$, the generalised $q$-coherent state turns into an ordinary $q$-coherent state,
\begin{align}
    \cket{\beta,\delta=0}=\sum_{n=0}^{\infty}\left(\frac{1-q}{\beta}-1\right)^{n}\frac{1}{\sqrt{(q;q)_n}}\ket{n}.
\end{align}
In particular, we have $\cket{1-q,0}=\ket{0}$.

Having said that, we can solve for the recurrence relation \eqref{eq:3term} generically.
We have
\begin{align}
    c_n = \left(i\sqrt{\frac{\delta}{\beta}}\right)^{n}\frac{H_n(X|q)}{\sqrt{(q;q)_n}}, \quad iX\equiv \frac{1-q-\beta+\delta}{2\sqrt{\beta\delta}},
\end{align}
which leads to
\begin{align}
    \braket{\theta|W}
    \langle\theta|W\rangle
    =
    \sqrt{\frac{(q,e^{\pm 2i\theta};q)_\infty}{2\pi}}
    \frac{(-\frac{\delta}{\beta};q)_\infty}
    {
    (\kappa_+(\beta,\delta)e^{\pm i\theta};q)_\infty
    (\kappa_-(\beta,\delta)e^{\pm i\theta};q)_\infty
    }.
    \label{eq:measure}
\end{align}
up to $\theta$-independent normalisation factors.

We further note a useful fact to be used later.
The state $q^{\ell \hat{N}}\cket{\beta,\delta}$ is proportional to another coherent state $\cket{\tilde{\beta},\tilde{\delta}}$ with different parameters;
we can prove that
\begin{align}
    \begin{split}
        \tilde{\beta}\equiv \frac{\beta}{\beta\left(\frac{1-q^\ell}{1-q}\right)+q^\ell\left(1+\delta\cdot\frac{1-q^\ell}{1-q}\right)}\\
        \tilde{\delta}\equiv \frac{\delta \cdot q^{2\ell}}{\beta\left(\frac{1-q^\ell}{1-q}\right)+q^\ell\left(1+\delta\cdot\frac{1-q^\ell}{1-q}\right)}
    \end{split},
    \label{eq:trickman}
\end{align}
or, quite simply,
\begin{align}
    \kappa_{\pm}(\tilde{\beta},\tilde{\delta})=q^{\ell}\kappa_{\pm}({\beta},{\delta}).
\end{align}

\subsection{$q\to 1$ limit}

\label{sec:defdef}

The limit $q\to 1$ is usually associated to some kind of continuous limit and will be interesting in our context later on.
Let us define $q\equiv e^{-\epsilon}$, so that the limit we are interested in is $\epsilon\to 0$.
In this limit, the operator $a^\dagger+a$ is known to become the Liouville Hamiltonian $\hat{D}_{\rm LQM}$.
More concretely, by rescaling our chord basis $\ket{n}$ by using
\begin{align}
    \phi\equiv \varphi+2\log \epsilon,\quad \varphi\equiv \epsilon n
\end{align}
we have \cite{Lin:2022rbf}
\begin{align}
    a^\dagger+a=\frac{2}{\sqrt{1-q}}-(1-q)^{3/2}\hat{D}_{\rm LQM}+O(\epsilon^{7/4}), \quad \hat{D}_{\rm LQM}\equiv -\frac{\mathrm{d}^2}{\mathrm{d}\phi^2}+e^{-\phi},
    \label{eq:defdef}
\end{align}
and
\begin{align}
    b^\dagger+b=\frac{2}{\sqrt{1-q}}-(1-q)^{3/2}\hat{D}_{\rm free}+O(\epsilon^{7/4}), \quad \hat{D}_{\rm free}\equiv -\frac{\mathrm{d}^2}{\mathrm{d}\xi^2},
    \label{eq:defdef111111}
\end{align}

Another interest is in how the coherent state $\cbra{\alpha,\gamma}$ and $\cket{\beta,\delta}$ behaves in the $q\to 1$ limit.
Our limit involves focusing on the $O(1/\epsilon)$-width window around $n_0\equiv 2\log \epsilon/\epsilon \gg 1$ in terms of $n$, and so we can safely approximate $c_n$ by using \eqref{eq:scaling} due to $n$ being large throughout the computation.
Then, along with some overall rescaling, we can replace them with 
\begin{align}
    \begin{split}
        \cket{\beta,\delta}\xrightarrow[\epsilon\to 0]{} \cket{u}\equiv \int_{-\infty}^{\infty} \mathrm{d}\phi\, e^{-u\phi}\ket{\phi},\quad \kappa_+(\beta,\delta)\equiv q^u,\\
        \cbra{\alpha,\gamma}\xrightarrow[\epsilon\to 0]{} \cbra{v}\equiv \int_{-\infty}^{\infty} \mathrm{d}\phi\, e^{-v\phi}\bra{\phi},\quad \kappa_+(\alpha,\gamma)\equiv q^v.
    \end{split}
    \label{eq:coherentlimit}
\end{align}
as $\epsilon\to 0$, with a normalising constant $\epsilon^2$ stripped off.
Note that the resulting states are not really normalisable even though $\cket{\beta,\delta}$ and $\cbra{\alpha,\gamma}$ are when $\kappa_+<1$.
We can think of the state as having a normalisation constant depending on $\epsilon$ in front, or we can make sure that the expression only makes sense inside correlators.

\section{ASEP and the open KPZ equation}
\label{sec:asep}


\subsection{ASEP and its stationary measure}

\subsubsection{Asymmetric simple exclusion process}

Asymmetric simple exclusion process (ASEP) is a continuous-time Markov process describing particles hopping on $N$ lattice sites aligned in one dimension.
As a stochastic process, each configuration of particles are assigned a probability;
We write a configuration of particles as $\vec{\uptau}\equiv (\uptau_1,\dots,\uptau_N)$, where $\uptau_i=1$ ($\uptau_i=0$) means a particle in present (absent) at site $i$, and write the probability of realising it as $p(\vec{\uptau})$.
It is customary to package probability distributions $p(\vec{\uptau})$ into a vector as
\begin{align}
    \ket{P}\equiv \sum_{\vec{\uptau}}p(\vec{\uptau})\ket{\vec{\uptau}},
\end{align}
where the time-evolution is governed by a Markov equation
\begin{align}
    \frac{\mathrm{d}}{\mathrm{d}t}\ket{P}=M\ket{P}.
\end{align}
The matrix $M$ is called the Markov matrix.
For later convenience, we denote the Hilbert space on site $i$ as $V_i$, which is isomorphic to $\mathbb{C}^2$.

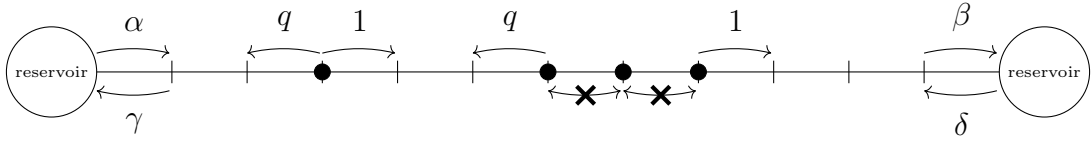
\begin{figure}
    \centering
    \begin{tikzpicture}[scale=1]
        \draw (-6.6,0) circle (0.6);
        \node at (-6.6,0) {{\tiny reservoir}};
        \draw (6.6,0) circle (0.6);
        \node at (6.6,0) {{\tiny reservoir}};
        \draw[->] (-6,0.25) arc (90+14:90-14:2);
        \node[above] at (-5.5,0.4) {{$\alpha$}};
        \draw[<-] (-6,-0.25) arc (-90-14:-90+14:2);
        \node[below] at (-5.5,-0.4) {{$\gamma$}};
        \draw (-6,0) -- (6,0); %
        \draw (-5,-0.15) -- (-5,0.15);
        \draw (-4,-0.15) -- (-4,0.15);
        \draw (-3,-0.15) -- (-3,0.15);
        \draw (-2,-0.15) -- (-2,0.15);
        \draw (-1,-0.15) -- (-1,0.15);
        \draw (0,-0.15) -- (0,0.15);
        \draw (1,-0.15) -- (1,0.15);
        \draw (2,-0.15) -- (2,0.15);
        \draw (3,-0.15) -- (3,0.15);
        \draw (4,-0.15) -- (4,0.15);
        \draw (5,-0.15) -- (5,0.15);
        \filldraw (-3,0) circle (3pt);
        \draw[->] (-3,0.25) arc (90+14:90-14:2);
        \node[above] at (-2.5,0.4) {{$1$}};
        \draw[<-] (-4,0.25) arc (90+14:90-14:2);
        \node[above] at (-3.5,0.4) {{$q$}};
        \filldraw (0,0) circle (3pt);
        \filldraw (1,0) circle (3pt);
        \draw[->] (2,0.25) arc (90+14:90-14:2);
        \node[above] at (2.5,0.4) {{$1$}};
        \draw[<-] (-1,0.25) arc (90+14:90-14:2);
        \node[above] at (-0.5,0.4) {{$q$}};
        \filldraw (2,0) circle (3pt);
        \draw[->] (5,0.25) arc (90+14:90-14:2);
        \node[above] at (5.5,0.4) {{$\beta$}};
        \draw[<-] (5,-0.25) arc (-90-14:-90+14:2);
        \node[below] at (5.5,-0.4) {{$\delta$}};
        \draw[<->] (1,-0.25) arc (-90-14:-90+14:2);
        \draw (1.5,-0.31) node[cross=5pt,ultra thick] {};
        \draw[<->] (0,-0.25) arc (-90-14:-90+14:2);
        \draw (0.5,-0.31) node[cross=5pt,ultra thick] {};
    \end{tikzpicture}
    \caption{A schematic picture indicating the update rules of ASEP.}
    \label{fig:asep}
\end{figure}

ASEP is defined \emph{via} the following update rules acting on particle configurations during infinitesimal time $\mathrm{d}t$:
\begin{itemize}
    \item A particle hops to the right at rate $1$ and left with rate $q$.
    \item Particles flow in at rate $\alpha$ and out at rate $\gamma$ on the left boundary.
    \item Particles flow out at rate $\beta$ and in at rate $\delta$ on the right boundary.
    \item Particles cannot hop to already occupied sites.
\end{itemize}
where something happening at rate $p$ really means happening at probability $p\,\mathrm{d}t$ during infinitesimal time $\mathrm{d}t$.
See also a schematic picture summarising the update rules in Figure \ref{fig:asep}.

We can also express the update rules in a Markov matrix as
\begin{align}
    M=
    \begin{pmatrix}
        -\alpha & \gamma  \\
        \alpha & -\gamma  \\
    \end{pmatrix}_{V_1}
    +
    \sum_{i=1}^{N-1}
    \begin{pmatrix}
        0 & 0 & 0 & 0 \\
        0 & -q & 1 & 0 \\
        0 & q & -1 & 0 \\
        0 & 0 & 0 & 0
    \end{pmatrix}_{V_i\otimes V_{i+1}}
    +
    \begin{pmatrix}
        -\delta & \beta  \\
        \delta & -\beta  \\
    \end{pmatrix}_{V_N},
\end{align}
where $M_1$, $M_N$, and $M_{V_i\otimes V_{i+1}}$ act as identity operators outside of $V_1$, $V_N$ and $V_{i}\otimes V_{i+1}$, respectively.
The bases of $V_i$ are given, from top to bottom columns and left to right rows, by $\ket{0}$ and $\ket{1}$, and of $V_{i}\otimes V_{i+1}$ by $\ket{0,0}$, $\ket{0,1}$, $\ket{1,0}$, and $\ket{1,1}$.

Let us also define the \emph{height function} for ASEP, which is crucial for relating ASEP to open KPZ equation later.
For a given configuration $\vec{\tau}$ the height function of ASEP is defined as
\begin{align}
    h_{\rm ASEP}(t,k)-h_{\rm ASEP}(t,0)\equiv \sum_{j=1}^{k}(2\tau_{j}-1),\quad \text{for integer $0 \leq k\leq N$}
\end{align}
where $h_{\rm ASEP}(0)$ is defined as $-2$ times the net number of particles which have entered though the left boundary at given time.
For non-integer values of $k$ the height function is defined \emph{via} linear interpolation.

\subsubsection{Stationary state of ASEP}
\label{sec:twolayer}

A finite Markov process is known to reach a steady state in the late-time limit where all the probabilities $p(\vec{\tau})$ become time-independent, under some assumptions.
ASEP is no exception, and hence it is an interesting question to ask what its stationary state is.
Utilising the underlying integrability, the ASEP stationary state was written down in terms of the following matrix product, \cite{derrida1993exact}
\begin{align}
    P(\vec{\tau})=\frac{\braket{W|\prod_{j=1}^{N}\left(\tau_j\DD + (1-\tau_j)\EE \right)|V}}{\braket{W|(\DD + \EE)^N|V}},
    \label{eq:probint}
\end{align}
where the matrices and the vectors satisfy the DEHP algebra, defined in \eqref{eq:DEHP}.
In particular, we have
\begin{align}
    \bra{W}\equiv \cbra{\alpha,\gamma}, \quad \ket{V}\equiv \cket{\beta,\delta}.
\end{align}
For later convenience, we define
\begin{align}
    \begin{gathered}
        A\equiv \kappa_+(\beta,\delta), \quad B\equiv \kappa_-(\beta,\delta),\quad
        C\equiv \kappa_+(\alpha,\gamma), \quad D\equiv \kappa_-(\alpha,\gamma),
    \end{gathered}
    \label{eq:ASEPparameters}
\end{align}
where $\kappa_{\pm}$ was defined in \eqref{eq:kappa}
and also
\begin{align}
    \rho_{\rm L}\equiv \frac{1}{1+C},\quad \rho_{\rm R}\equiv \frac{A}{1+A}.
\end{align}
The intuitive meaning of $\rho_{\rm L,\,R}$ is the densities of fictitious particles on site $i=0,\, N+1$, respectively.

Note that for \eqref{eq:probint} to make sense probabilistically, we need a condition $\braket{W|V}<\infty$.
This is translated to $AC<1$ or equivalently $\rho_{\rm L}>\rho_{\rm R}$.
The region is called the fan region in the ASEP context.
Such an assumption can be relaxed by going to a different representation of DEHP algebra \cite{uchiyama2004asymmetric,bryc2017asymmetric,wang2024askey}, which we also discuss in Appendix \ref{sec:fin}.

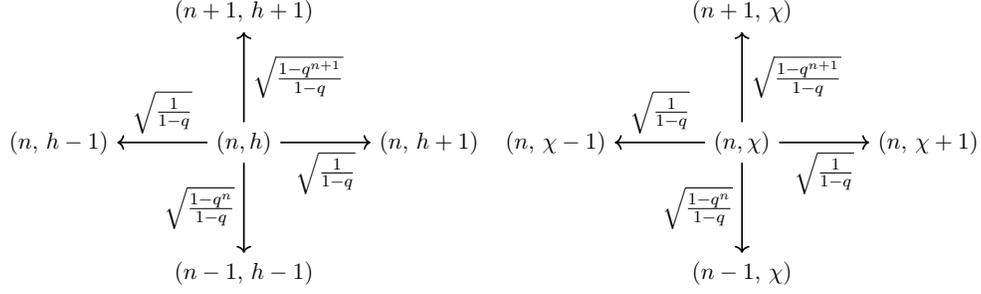
\begin{figure}[t]
    \centering
    \subfloat{\scalebox{0.8}{\begin{tikzpicture}[every node/.style={font=\small},node distance=1.5cm]
        \node (C) {$(n,h)$};
        \node[right =of C]  (E) {$(n,\,h+1)$};
        \node[above =of C]  (N) {$(n+1,\,h+1)$};
        \node[left  =of C]  (W) {$(n,\,h-1)$};
        \node[below =of C]  (S) {$(n-1,\,h-1)$};
        \draw[->, thick] (C) -- node[midway, below] {$\sqrt{\frac{1}{1-q}}$} (E);
        \draw[->, thick] (C) -- node[midway, right] {$\sqrt{\frac{1-q^{n+1}}{1-q}}$} (N);
        \draw[->, thick] (C) -- node[midway, above] {$\sqrt{\frac{1}{1-q}}$} (W);
        \draw[->, thick] (C) -- node[midway, left]  {$\sqrt{\frac{1-q^{n}}{1-q}}$} (S);
    \end{tikzpicture}}
    }
    \subfloat{\scalebox{0.8}{\begin{tikzpicture}[every node/.style={font=\small},node distance=1.5cm]
        \node (C) {$(n,\chi)$};
        \node[right =of C]  (E) {$(n,\,\chi+1)$};
        \node[above =of C]  (N) {$(n+1,\,\chi)$};
        \node[left  =of C]  (W) {$(n,\,\chi-1)$};
        \node[below =of C]  (S) {$(n-1,\,\chi)$};
        \draw[->, thick] (C) -- node[midway, below] {$\sqrt{\frac{1}{1-q}}$} (E);
        \draw[->, thick] (C) -- node[midway, right] {$\sqrt{\frac{1-q^{n+1}}{1-q}}$} (N);
        \draw[->, thick] (C) -- node[midway, above] {$\sqrt{\frac{1}{1-q}}$} (W);
        \draw[->, thick] (C) -- node[midway, left]  {$\sqrt{\frac{1-q^{n}}{1-q}}$} (S);
    \end{tikzpicture}}
    }
    \caption{(Left) A random walk update rule governing the evolution of the ASEP height function, with $n$ being an auxiliary coordinate. (Right) A reparametrisation of the left update rule using $\chi\equiv h-n$.}
    \label{fig:randomwalk}
\end{figure}

Let us interpret \eqref{eq:probint} so that we have a better understanding of the height function and the chord number basis.
The discussion closely follows that  of \cite{barraquand2023stationary}, except that we use a different basis amenable to relating it to double-scaled SYK.
Given a partial configuration $\vec{\tau}$, associated is the probability distribution $P(\tau)$ (or equivalently, $P(\vec{h}_{\rm ASEP})$, for the height function configuration).
Inserting a complete set at each step in \eqref{eq:probint}, we can break $P(\tau)$ into a sum,
\begin{align}
    P(\vec{\tau})=\frac{1}{Z_N} \sum_{\vec{n}}\braket{V|n_0}\braket{n_0|\mathsf{X}_1|n_1}\cdots \braket{n_{N-1}|\mathsf{X}_N|n_N}\braket{n_N|W},
    \label{eq:crucial}
\end{align}
where we define $Z_N\equiv \braket{V|(\DD+\EE)^N|W}$ and
\begin{align}
    \mathsf{X}_i = \frac{1}{\sqrt{1-q}}+\mathsf{x}_i,
    \quad 
    \mathsf{x}_i\equiv  
    \begin{cases}
        a & \text{when $h_i-h_{i-1}=1$} \\
        a^\dagger & \text{when $h_i-h_{i-1}=-1$}
    \end{cases},
\end{align}
This makes it possible to interpret each summand as a joint probability of realising the configuration $\vec{n}$ \emph{and} $\vec{\tau}$ (or again, equivalently $\vec{h}_{\rm ASEP}$):
\begin{align}
    p(\vec{n},\vec{h}_{\rm ASEP})\equiv \frac{1}{Z_N}\braket{V|n_0}\braket{n_0|\mathsf{X}_1|n_1}\cdots \braket{n_{N-1}|\mathsf{X}_N|n_N}\braket{n_N|W}.
\end{align}
Furthermore, because of the product structure, the probability distribution can be understood as a Markov process, where the state takes value in $(n,h)\in \mathbb{Z}^2$, with $n\geq 0$.
In other words, $\braket{n_{i-1}|\mathsf{X}_i|n_i}$ gives an unnormalised probability of hopping from $(n_{i-1},h_{i-1})$ to $(n_{i},h_{i})$, see the left panel of Figure \ref{fig:randomwalk} for the schematically shown update rule.
In this interpretation, $\braket{V|n_0}$ and $\braket{n_N|W}$ give the initial and the final probability distribution, respectively.
Interestingly, with linear reparametrisation, we can represent $h_{\rm ASEP}=n+\chi$, where $\chi$ evolves as an unbiased simple random walk.
Put differently, the evolution of the height function $h_{\rm ASEP}$ is governed by an underlying two-dimensional random walk $(n,\chi)$, with an identification that $h_{\rm ASEP}=n+\chi$.
See the right panel of Figure \ref{fig:randomwalk} for the schematically shown update rule for the newly parametrised random walk.

To sum up, the stationary measure of the open ASEP is described by a transfer matrix $\mathsf{T}_{\rm ASEP}$, given by
\begin{align}
    \mathsf{T}_{\rm ASEP}\equiv a^\dagger \otimes \mathbbm{1} + a \otimes \mathbbm{1} + \mathbbm{1}\otimes b^\dagger + \mathbbm{1}\otimes b,
\end{align}
with creation/annihilation operators already defined in \eqref{eq:aaa} and in \eqref{eq:bbb}.
We also redefine $\bra{W}$ and $\ket{V}$ to give a homogeneous distribution in $\chi$, in other words, $\bra{W}\otimes \sum_{\chi\in\mathbb{Z}}\bra{\chi}$ and $\ket{V}\otimes \sum_{\chi\in\mathbb{Z}}\ket{\chi}$.
Importantly, any relevant physical quantities can be computed using this transfer matrix.
For example, the probability of realising $h_i$ at step $i$ is simply given by
\begin{align}
    p(h_i)=\frac{1}{Z_N}\sum_{n=0}^{\infty}\braket{W|({\mathsf{T}_{\rm ASEP}})^i|n,h_i-n}\braket{n,h_i-n|({\mathsf{T}_{\rm ASEP}})^{N-i}|V}.
\end{align}

\subsection{Weakly asymmetric limit to open KPZ}

\subsubsection{Weakly asymmetric limit}

Kardar--Parisi--Zhang (KPZ) equation is a stochastic differential equation describing a broad class of driven interfaces.
It is an stochastic evolution equation of a height profile $h(t,x)$, where $t$ is time and $x$ is the one-dimensional spatial coordinate; 
It is written as
\begin{align}
\partial_t h
= {\partial_x^2}h +\frac{1}{2} \left(\partial_x h\right)^2 + \xi(x,t).
\end{align}
where $\xi$ is a spacetime white noise satisfying $\braket{\xi(t,x)\xi(t^\prime,x^\prime)}=\delta(t-t^\prime)\delta(x-x^\prime)$.
Note that, by rescaling units, we set all the dimensionful constants to one.

We are particularly interested in the open KPZ equation, which is defined on an interval $0\leq x\leq X$.
The boundary condition we impose is the Neumann boundary condition,\footnote{Some care is needed to properly define such boundary conditions because $h(t,x)$ is not differentiable, but we will not discuss this subtlety any further.} where 
\begin{align}
    \partial_xh(t,x)\bigr|_{x=0}=u, \quad \partial_xh(t,x)\bigr|_{x=X}=-v.
    \label{eq:Neumann}
\end{align}
Our interest lies in the invariant measure for the open KPZ equation, in other words its time-independent solution.

The open KPZ equation is known to be obtained as a certain scaling limit of ASEP called the weakly asymmetric limit.
In \cite{corwin2018open,parekh2019kpz,yang2025kpz}, it was proven that the rescaled ASEP height function
\begin{align}
    h^{(N)}(t,x)\equiv \frac{1}{\sqrt{N_{\rm d}}}h_{\rm ASEP}\left(\frac{1}{2}q^{-1/2}N_{\rm d}^2 t,\lfloor{{N_{\rm d}}x}\rfloor\right)+\left(\frac{1}{2N_{\rm d}}+\frac{1}{24}\right)t,
    \label{eq:waseplimit}
\end{align}
where $N_{\rm d}\equiv \frac{N}{X}$,
converges to the height function of the open KPZ equation (\emph{i.e.}, satisfies the open KPZ equation) in the $N\to\infty$ limit, with parameters scaling as
\begin{align}
    q=\exp\left(-\frac{2}{\sqrt{N_{\rm d}}}\right)
\end{align}
and 
\begin{align}
    \begin{gathered}
        \alpha = \frac{1-q}{(1-q^r)(1+q^u)}, \quad \beta=\frac{1-q}{(1-q^s)(1+q^v)},\\
        \gamma = \frac{q^{u+r}(1-q)}{(1-q^r)(1+q^u)}, \quad \delta=\frac{q^{v+s}(1-q)}{(1-q^s)(1+q^v)}.
    \end{gathered}
\end{align}
Here $u$ and $v$ parametrise the Neumann boundary condition as in \eqref{eq:Neumann}, whereas $s,r>0$ are arbitrary parameters.
The choice of parameters is equivalent to setting
\begin{align}
    A=q^v,\quad B=-q^s,\quad C=q^u,\quad D=-q^r.
\end{align}
This means that we are zooming into the triple point in the ASEP phase diagram, $\rho_{\rm L}=\rho_{\rm R}=1/2$ in the $N\to \infty$ limit.

\subsection{Open KPZ stationary measure}

The weakly asymmetric limit presented in the previous subsection is in fact exactly the same as the limit defined in Section \ref{sec:defdef}, with the identification of $\epsilon=2/\sqrt{N_{\rm d}}$.
To see this, notice the fact that $h_{\rm ASEP}$ should be rescaled with $1/\sqrt{N_{\rm d}}$; this suggests that we rescale $n$ and $\chi$ in the same way, because $h_{\rm ASEP}=n+\chi$.
In light of this, we define $\phi\equiv n/\sqrt{N_{\rm d}}+2\log (1/\sqrt{N_{\rm d}})$ and $\xi\equiv \chi/\sqrt{N_{\rm d}}$, which is indeed the limit taken in Section \ref{sec:defdef}, after rescaling the lattice constant as $1/N_{\rm d}$.

We can then use this fact to study the path-integral measure for $h$ as $x$ ranges from $0$ to $X$.
By using \eqref{eq:defdef} and \eqref{eq:defdef111111}, we see that
the transition amplitude for $h$ from $x$ to $x+\delta x$ is given by
\begin{align}
    (\mathsf{T}_{\rm ASEP})^{N_{\rm d}\cdot\delta x}\propto \left(1-\frac{1}{N_{\rm d}}(\hat{D}_{\rm LQM}+\hat{D}_{\rm free})\right)^{N_{\rm d}\cdot\delta x}
    \xrightarrow[N_{\rm d}\to \infty]{}e^{-\delta x (\hat{D}_{\rm LQM}+\hat{D}_{\rm free})}.
    \label{eq:wert}
\end{align}
This means that we are taking a continuous limit of the random walk described by $(n,\chi)$ to get a Euclidean evolution of states described by a sum of Liouville quantum mechanics and a free scalar.
In other words, the Hamiltonian that generates the stationary configuration for $h$ can be written as
\begin{align}
    \hat{D}_{\rm JT}\equiv \hat{D}_{\rm LQM}+\hat{D}_{\rm free}=-\frac{\mathrm{d}^2}{\mathrm{d}\phi^2}+e^{-\phi}-\frac{\mathrm{d}^2}{\mathrm{d}\xi^2}.
\end{align}

Likewise, in the weakly asymmetric limit, the states at the edge of the interval where KPZ equation become
\begin{align}
    \begin{split}
        \ket{W}\xrightarrow[\epsilon\to 0]{} \cket{u}\equiv \int_{-\infty}^{\infty} \mathrm{d}\phi\, e^{-u\phi}\ket{\phi},\quad \kappa_+(\beta,\delta)\equiv q^u,\\
        \bra{V}\xrightarrow[\epsilon\to 0]{} \cbra{v}\equiv \int_{-\infty}^{\infty} \mathrm{d}\phi\, e^{-v\phi}\bra{\phi},\quad \kappa_+(\alpha,\gamma)\equiv q^v,
    \end{split}
    \label{eq:coherentlimitASEP}
\end{align}
which just copies \eqref{eq:coherentlimit}.
As in the case of ASEP, we take $\cket{v}$ and $\cbra{u}$ to have a homogeneous distribution in $\xi$ with a slight abuse of notation, \emph{i.e.}, $\cket{v}\otimes \int d\xi \ket{\xi}$ and $\cbra{u}\otimes \int d\xi \bra{\xi}$.

To get physical quantities out of these, we can simply follow the ASEP case.
For example, to get a probability of realising a stationary configuration satisfying $h(x_0)=h_0$ at a specific point $x=x_0$,
we compute
\begin{align}
    p[h(x_0)=h_0]=\frac{1}{Z(X)}\cbraket{u|e^{-x_0(\hat{D}_{\rm LQM}+\hat{D}_{\rm free})}\mathbb{P}_{h_0}e^{-(X-x_0)\cdot(\hat{D}_{\rm LQM}+\hat{D}_{\rm free})}|v}
\end{align}
where $\mathbb{P}_{h_0}$ is a projector onto a subspace satisfying $\phi+\xi=h_0$.
We have also implicitly redefined $\cbra{u}$ and $\cket{v}$, originally defined in \eqref{eq:coherentlimit}, to be homogeneous in the $\xi$-direction. 
Note also that the overall normalisation is defined as
\begin{align}
    Z(X)\equiv \cbraket{u|e^{-X(\hat{D}_{\rm LQM}+\hat{D}_{\rm free})}|v}.
\end{align}

\section{Double-scaled SYK and JT gravity}
\label{sec:dssyk}


\subsection{Double-scaled SYK}

The Sachdev--Ye--Kitaev (SYK) model is a quantum mechanical model of $\mathtt{N}$ Majorana fermions with all-to-all random $p$-body interactions.
Denoting the $\mathtt{N}$ fermions as $\psi_i$ ($i=1,\,2,\,\dots,\, \mathtt{N}$) with anti-commutation relations $\{\psi_i,\psi_j\}=2\delta_{ij}$, the Hamiltonian of the SYK model reads
\begin{align} \label{eq:syk}
H = i^{{\mathtt{p}}/2} \sum _{1 \le i_1<\cdots <i_{{\mathtt{p}}}\le \mathtt{N} } J_{i_1 i_2 \cdots i_{\mathtt{p}}} \psi_{i_1} \cdots \psi_{i_{\mathtt{p}}}.
\end{align}
Here we take $J_{i_1 i_2 \cdots i_{\mathtt{p}}}$ to be Gaussian distributed with variance
\begin{align}
    \left\langle J_{i_1 i_2 \cdots i_{\mathtt{p}}}J_{j_1 j_2 \cdots j_{\mathtt{p}}}\right\rangle={\binom{\mathtt{N}}{{\mathtt{p}}}}^{-1}\delta_{i_1,j_1}\cdots \delta_{i_{\mathtt{p}},j_{\mathtt{p}}},
    \label{eq:variance}
\end{align}
which is tantamount to setting $\left\langle\Tr[H^2]\right\rangle=1$, with $\Tr$ denoting a normalised trace characterised by $\Tr[\mathbbm{1}]=1$. 
Throughout the paper, the bracket $\braket{\cdot}$ denotes taking an average over these random couplings.

The SYK model has an interesting solvable limit aside from the usual large-$\mathtt{N}$, fixed-$\mathtt{p}$ limit.
The limit we are interested in is the double-scaling limit where we take
\begin{align}
    \mathtt{N}\to\infty,\quad \lambda\equiv \frac{2{\mathtt{p}}^2}{\mathtt{N}}=\fixed, \quad \mathtt{q}\equiv e^{-\lambda},
\end{align}
and the resulting model is called the double-scaled SYK.

\subsection{Chord diagrams}

\subsubsection{Transfer matrix on chord Hilbert space}

The double-scaled SYK can be solved using a combinatorial tool called the chord diagram.
In particular, a crucial observation in \cite{Berkooz:2018jqr} is that the physical quantities in double-scaled SYK can be computed by using a transfer matrix on the auxiliary Hilbert space, called the \emph{chord Hilbert space}. 
The readers are referred to \cite{Berkooz:2024lgq} for reviews.

Let us quickly see how we arrive at the chord Hilbert space.
Imagine computing the thermal partition function $\Braket{\Tr[e^{-\beta H}]}$ by computing the
$k$-th thermal moment $m_k\equiv \Braket{\Tr[H^k]}$.
There are in total $k$ Gaussian random variables $J_{i_1\cdots i_{\mathtt{p}}}$ in $m_k$, and so when we take an average over them, we make pairs and contract the indices in each pair.
Each such contraction can be represented as a circle with $k$ sites with lines pairing them up, \emph{i.e.}, a \emph{chord diagram} (Table \ref{tab:chord}). 
A combinatorial argument then assigns a value to a chord diagram as $\mathtt{q}^{\mathtt{\#intersections}}$, where $\mathtt{\#intersections}$ is the number of chord intersections in a give chord diagram.
We therefore end up with a formula
\begin{align}
    m_k=\sum_{\text{chord diagrams}} \mathtt{q}^{\mathtt{\#intersections}}.
\end{align}
We depict an example of chord diagrams for $k=4$ in Table \ref{tab:chord}.

\begin{table}[t]
    \centering
    \begin{tabular}{c|c|c}
        diagram & expression & value \\ \hline\hline
        \begin{tikzpicture}[scale=0.3,baseline=(O.base)]
            \node (O) {};
            \draw (0,0) circle (2cm); %
            \filldraw (45:2cm) circle (5pt); %
            \node[above right] at (45:2cm) {\footnotesize $I_4$};
            \filldraw (135:2cm) circle (5pt); %
            \node[above left] at (135:2cm) {\footnotesize $I_1$};
            \filldraw (225:2cm) circle (5pt); %
            \node[below left] at (225:2cm) {\footnotesize $I_2$};
            \filldraw (315:2cm) circle (5pt); %
            \node[below right] at (315:2cm) {\footnotesize $I_3$};
            \draw (135:2cm) -- (225:2cm); %
            \draw (45:2cm) -- (315:2cm);
            \end{tikzpicture} & $\displaystyle i^{2\mathtt{p}}\sum_{I_1,I_2,I_3,I_4}\braket{
        J_{I_1}J_{I_2}}\braket{J_{I_3} J_{I_4}}\Tr\left[\psi_{I_1}\psi_{I_2}\psi_{I_3} \psi_{I_4}\right]$ & $1$
        \\ \hline
        {\begin{tikzpicture}[scale=0.3,baseline=(O.base)]
        \node (O) {};
            \draw (0,0) circle (2cm); %
            \filldraw (45:2cm) circle (5pt); %
            \node[above right] at (45:2cm) {\footnotesize $I_4$};
            \filldraw (135:2cm) circle (5pt); %
            \node[above left] at (135:2cm) {\footnotesize $I_1$};
            \filldraw (225:2cm) circle (5pt); %
            \node[below left] at (225:2cm) {\footnotesize $I_2$};
            \filldraw (315:2cm) circle (5pt); %
            \node[below right] at (315:2cm) {\footnotesize $I_3$};
            \draw (45:2cm) -- (225:2cm); %
            \draw (135:2cm) -- (315:2cm);
            \end{tikzpicture}}
            &
            $\displaystyle i^{2\mathtt{p}}\sum_{I_1,I_2,I_3,I_4}\braket{
        J_{I_1}J_{I_3}}\braket{J_{I_2} J_{I_4}}\Tr\left[\psi_{I_1}\psi_{I_2}\psi_{I_3} \psi_{I_4}\right]$ & $\mathtt{q}$\\ \hline
        {\begin{tikzpicture}[scale=0.3,baseline=(O.base)]
        \node (O) {};
            \draw (0,0) circle (2cm); %
            \filldraw (45:2cm) circle (5pt); %
            \node[above right] at (45:2cm) {\footnotesize $I_4$};
            \filldraw (135:2cm) circle (5pt); %
            \node[above left] at (135:2cm) {\footnotesize $I_1$};
            \filldraw (225:2cm) circle (5pt); %
            \node[below left] at (225:2cm) {\footnotesize $I_2$};
            \filldraw (315:2cm) circle (5pt); %
            \node[below right] at (315:2cm) {\footnotesize $I_3$};
            \draw (45:2cm) -- (135:2cm); %
            \draw (225:2cm) -- (315:2cm);
            \end{tikzpicture}}
            &
            $\displaystyle i^{2\mathtt{p}}\sum_{I_1,I_2,I_3,I_4}\braket{
        J_{I_1}J_{I_4}}\braket{J_{I_2} J_{I_3}}\Tr\left[\psi_{I_1}\psi_{I_2}\psi_{I_3} \psi_{I_4}\right]$ & $1$
    \end{tabular}
    \caption{A list of chord diagrams appearing in $m_4$, and corresponding expressions and their values. $I$ denotes an ordered index set collecting $i_1\cdots i_{\mathtt{p}}$. We see that $m_4$ evaluates to $m_4=2+\mathtt{q}$.} 
    \label{tab:chord}
\end{table}

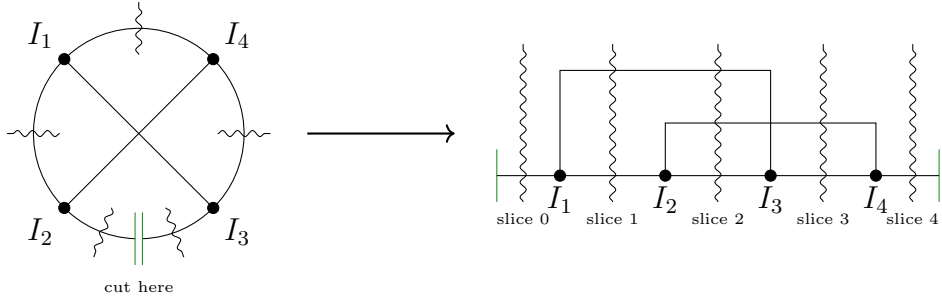
\begin{figure}[t]
    \centering
    \begin{tikzpicture}[scale=0.7]
        \filldraw (45:2cm) circle (3pt); %
        \node[above right] at (45:2cm) {$I_4$};
        \filldraw (135:2cm) circle (3pt); %
        \node[above left] at (135:2cm) {$I_1$};
        \filldraw (225:2cm) circle (3pt); %
        \node[below left] at (225:2cm) {$I_2$};
        \filldraw (315:2cm) circle (3pt); %
        \node[below right] at (315:2cm) {$I_3$};
        \draw (45:2cm) -- (225:2cm); %
        \draw (135:2cm) -- (315:2cm);
        \draw (268:2cm) arc (268:-88:2cm);
        \draw [decorate, decoration={snake, amplitude=.4mm, segment length=2mm}] (250:1.5cm) -- (250:2.5cm);
        \draw [decorate, decoration={snake, amplitude=.4mm, segment length=2mm}] (0:1.5cm) -- (0:2.5cm);
        \draw [decorate, decoration={snake, amplitude=.4mm, segment length=2mm}] (90:1.5cm) -- (90:2.5cm);
        \draw [decorate, decoration={snake, amplitude=.4mm, segment length=2mm}] (290:1.5cm) -- (290:2.5cm);
        \draw [decorate, decoration={snake, amplitude=.4mm, segment length=2mm}] (180:1.5cm) -- (180:2.5cm);
        \draw[ForestGreen] (268:2cm) ++(0, 0.5) -- ++(0, -1); %
        \draw[ForestGreen] (272:2cm) ++(0, 0.5) -- ++(0, -1);  %
        \node[below] at (270:2.6cm) {{\tiny cut here}};
        \draw[->, thick, black] (3.2, 0) -- (6, 0);
        \begin{scope}[shift={(11,-0.8)}]
            \draw[ForestGreen] (-4.2,0) ++(0, 0.5) -- ++(0, -1); %
            \draw[ForestGreen] (4.2,0) ++(0, 0.5) -- ++(0, -1);  %
            \draw (-4.2,0) -- (4.2,0); %
            \draw (-3,0) -- (-3,2) -- (1,2) -- (1,0); %
            \draw (-1,0) -- (-1,1) -- (3,1) -- (3,0); %
            \draw [decorate, decoration={snake, amplitude=.4mm, segment length=2mm}] (-3.7,-0.5) -- (-3.7,2.5);
            \node[below] at (-3.7,-0.5) {{\tiny slice 0}};
            \filldraw (-3,0) circle (3pt); %
            \node[below] at (-3,0) {$I_1$};
            \draw [decorate, decoration={snake, amplitude=.4mm, segment length=2mm}] (-2,-0.5) -- (-2,2.5);
            \node[below] at (-2,-0.5) {{\tiny slice 1}};
            \filldraw (-1,0) circle (3pt); %
            \node[below] at (-1,0) {$I_2$};
            \draw [decorate, decoration={snake, amplitude=.4mm, segment length=2mm}] (-0,-0.5) -- (-0,2.5);
            \node[below] at (-0,-0.5) {{\tiny slice 2}};
            \filldraw (1,0) circle (3pt); %
            \node[below] at (1,0) {$I_3$};
            \draw [decorate, decoration={snake, amplitude=.4mm, segment length=2mm}] (2,-0.5) -- (2,2.5);
            \node[below] at (2,-0.5) {{\tiny slice 3}};
            \filldraw (3,0) circle (3pt); %
            \node[below] at (3,0) {$I_4$};
            \draw [decorate, decoration={snake, amplitude=.4mm, segment length=2mm}] (3.7,-0.5) -- (3.7,2.5);
            \node[below] at (3.7,-0.5) {{\tiny slice 4}};
        \end{scope}
    \end{tikzpicture}
    \caption{A procedure of cutting open a chord diagram. The wiggly lines represent the slices on which the chord Hilbert space is defined.
    }
    \label{fig:opening}
\end{figure}

Although the computation of $m_k$ by simple enumeration goes quickly out of hand as we increase $k$,
there is a way to compute it, called the transfer matrix method.
The transfer matrix tracks the number of intersections as we move from one site to the other in the chord diagram, after we cut it open (Figure \ref{fig:opening}).
By using such a technique, we are able to express the $k$-th thermal moment as 
\begin{align}
    m_k = \braket{0|T^k|0},
\end{align}
where we let $T$ the transfer matrix.
The chord Hilbert space on which $T$ acts is spanned by the basis $\ket{n}$, which is a state representing the configuration where there are $n$ chords on the slice in question.
In this way of counting chord intersections, the quantity $\braket{n|T^i|0}$, for example, gives us the sum over the set of chord diagrams, weighed by $\mathtt{q}^{\#\mathtt{intersections}}$, which has $n$ chords at step $i$.

The form of the transfer matrix is given by the following:
\begin{align}
    T\ket{n}\equiv \sqrt{\frac{1-\mathtt{q}^{n+1}}{1-\mathtt{q}}}\ket{n+1}+\sqrt{\frac{1-\mathtt{q}^n}{1-\mathtt{q}}}\ket{n-1}=(a^\dagger+a)\ket{n},
\end{align}
where $a^\dagger$ and $a$ are defined in \eqref{eq:aaa}.
We refer the reader again to \cite{Berkooz:2024lgq} for derivations; note that with $T$ being a tri-diagonal matrix, we have already symmetrised it \emph{via} a change of basis.
The transfer matrix can be viewed as either opening or closing a chord with \revisionaddition{transfer weight} $\sqrt{\frac{1-\mathtt{q}^{n+1}}{1-\mathtt{q}}}$ or $\sqrt{\frac{1-\mathtt{q}^{n}}{1-\mathtt{q}}}$, respectively, after a \revisionaddition{chord-counting step}.

The transfer matrix can be diagonalised by using \eqref{eq:qqqqqqq}.
We have
\begin{align}
    T\ket{\theta}&=\frac{2\cos\theta}{\sqrt{1-\mathtt{q}}}\ket{\theta},
\end{align}
which leads to the computation of the thermal partition function as
\begin{align}
    \braket{\Tr[e^{-\beta H}]} = \braket{0|e^{-\beta T}|0}=\int_0^\pi \frac{d\theta}{2\pi}\,(\mathtt{q},e^{\pm 2i\theta};\mathtt{q})_\infty \exp\left(-\beta\frac{2\cos\theta}{\sqrt{1-\mathtt{q}}}\right),
\end{align}
by using \eqref{eq:coordinate}
We also see that the vacuum energy $E_0$ of $T$ is given at $\theta =\pi$ to be $E_0=-\frac{2}{\sqrt{1-\mathtt{q}}}$.

In the following, we will be more interested in the vacuum-subtracted version of $T$ rather than $T$ itself.
We define
\begin{align}
    \mathsf{T}_{\rm DSSYK}^{\mathcal{N}=0}\equiv T+\frac{2}{\sqrt{1-\mathtt{q}}}=\DD+\EE, \quad \mathsf{H}\equiv H+\frac{2}{\sqrt{1-\mathtt{q}}}
\end{align}
where $E_0\equiv -\frac{2}{\sqrt{1-\mathtt{q}}}$ is the ground state energy of the double-scaled SYK model.

\subsubsection{Doubled Hilbert space}

An attractive interpretation of the chord Hilbert space is that it describes the dynamics of two decoupled SYK models, rather than one \cite{Lin:2022rbf}.
To motivate this, let us rewrite  
\begin{align}
    {\Tr[\mathsf{H}^k]}=\braket{\Omega|{\mathsf{H}_L}^i{\mathsf{H}_R}^{k-i}|\Omega} \quad \text{for any $i$}
\end{align}
by using the maximally entangled state $\ket{\Omega}$ between the two double-scaled SYK Hilbert spaces, and 
$\mathsf{H}_L$ and $\mathsf{H}_R$ denote the left and the right Hamiltonians, respectively.
We can compare this with the chord expression $\braket{0|\mathsf{T}^k|0}$; this motivates us to think of the chord state $\ket{0}$ as corresponding to $\ket{\Omega}$, whereas the chord transfer matrix $\mathsf{T}_{\rm DSSYK}^{\mathcal{N}=0}$ as $\mathsf{T}_L$ or $\mathsf{T}_R$.
(There are no distinctions between $\mathsf{H}_L$ and $\mathsf{H}_R$ as they act the same way on $\ket{\Omega}$.)

\begin{figure}[t]
    \centering
    \begin{tikzpicture}[scale=1]
        \draw (263:2cm) arc (263:97:2cm);
        \draw (83:2cm) arc (83:-83:2cm);
        \draw [decorate, decoration={snake, amplitude=.4mm, segment length=2mm}] (66:2.5cm) arc (0:-180:1cm);
        \draw [decorate, decoration={snake, amplitude=.4mm, segment length=2mm}] (-66:2.5cm) arc (0:180:1cm);
        \draw [decorate, decoration={snake, amplitude=.4mm, segment length=2mm}] (0:2.5cm) -- (180:2.5cm);
        \draw [decorate, decoration={snake, amplitude=.4mm, segment length=2mm}] (30:2.4cm) arc (-53:-127:3.5cm);
        \draw [decorate, decoration={snake, amplitude=.4mm, segment length=2mm}] (-30:2.4cm) arc (53:127:3.5cm);
        \draw [ForestGreen] (-83:2cm) arc (0:180:0.25cm);
        \draw [ForestGreen] (83:2cm) arc (0:-180:0.25cm);
        \node at (90:2.6cm) {\rotatebox{-90}{\large $\langle 0 \!\mid$}};
        \node at (-90:2.6cm) {\rotatebox{-90}{\large $\mid\! 0 \rangle$}};
        \filldraw (45:2cm) circle (3pt); %
        \filldraw (105:2cm) circle (3pt); %
        \filldraw (165:2cm) circle (3pt); %
        \filldraw (225:2cm) circle (3pt); %
        \filldraw (285:2cm) circle (3pt); %
        \filldraw (345:2cm) circle (3pt); %
    \end{tikzpicture}
    \caption{An interpretation of the chord Hilbert space.
    We start and end with a chord vacuum state corresponding to a maximally entangled state in the double-scaled SYK.
    The wiggly lines are slices on which our chord states are defined.
    They evolve according to the transfer matrices (represented as sites) acting on the left and the right Hilbert spaces, from/on which chords can emanate/close. 
    }
    \label{fig:opening1}
\end{figure}
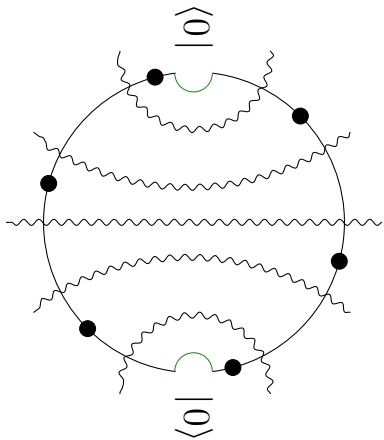

Pushing forward such an analogy leads us to the following picture (Figure \ref{fig:opening1}) \cite{Lin:2022rbf}.
We cut open the chord diagram at antipodal points, and interpret the left and the right arcs as the left and the right SYK model.
The state living on the antipodal points are the chord state $\ket{0}$, which corresponds to the maximally entangled state of the original theory.
The chord states then can be thought of as living on the wiggly lines in Figure \ref{fig:opening1}, whereas the sites represent the action of the transfer matrix.
In this picture, the chord state $\ket{n}$ on a wiggly line corresponds to the situation where there are $n$ open chords on the line.
What is important is that there is no need to distinguish between $\mathsf{T}_L$ and $\mathsf{T}_R$ whatsoever here -- It makes no difference on which side you open/close a chord.

\subsection{End-of-the-world branes}

\begin{figure}[t]
    \centering
    \begin{tikzpicture}[scale=1]
        \draw (225:2cm) arc (225:116:2cm);
        \draw (64:2cm) arc (64:-45:2cm);
        \draw [ultra thick, BlueViolet] (64:2cm) arc (-26:-154:1cm);
        \draw [ultra thick, BrickRed] (-45:2cm) arc (55:125:2.5cm);
        \node at (90:2.1cm) {\rotatebox{-90}{\large $\langle {V} \!\mid$}};
        \node at (-90:1.8cm) {\rotatebox{-90}{\large $\mid\! {W} \rangle$}};
        \filldraw (45:2cm) circle (3pt); %
        \filldraw (165:2cm) circle (3pt); %
        \filldraw (345:2cm) circle (3pt); %
    \end{tikzpicture}
    \caption{A chord representation of the amplitude between two EOW branes as in \eqref{eq:form2ndssyk2eow}, in this case $\braket{V|(\mathsf{T}_{\rm DSSYK}^{\mathcal{N}=0})^3|W}$. As in Figure \ref{fig:opening1}, the state evolves according to the transfer matrices (represented as sites) acting on the left and the right Hilbert spaces.}
    \label{fig:eowfig}
\end{figure}
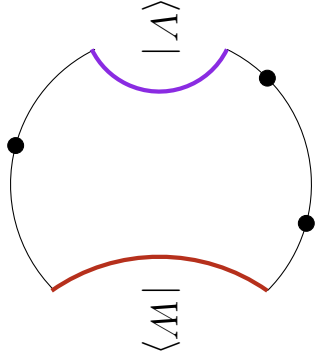

We have been considering the vacuum-to-vacuum amplitude in double-scaled SYK model, but there are other interesting amplitudes that generalises it.
Of particular interest to us are the amplitude between two end-of-the-world (EOW) branes, $\bra{{V}}\equiv \cbra{\alpha,\gamma}$ and $\ket{{W}}\equiv \cket{\beta,\delta}$, which are a slight generalisations of \cite{Okuyama:2023byh}.
In \cite{Okuyama:2023byh}, a special case with $\gamma=\delta=0$ has been discussed, and it was argued that $(q,e^{2\pm i\theta};q)\braket{\theta|V}$ has been argued to be the orthogonality measure for the continuous big $q$-Hermite polynomials.
In our case, we have \eqref{eq:measure}, which means that $(q,e^{2\pm i\theta};q)\braket{\theta|V}$ is the orthogonality measure for the Al-Salam--Chihara polynomials.
As discussed in \cite{Berkooz:2025ydg}, it corresponds to the fact that the transfer matrix can be diagonalised using Al-Salam--Chihara polynomials when we consider the time-evolution with one end of the time-slice ending on the ordinary boundary of the chord diagram, while the other of the reservoir region representing the EOW brane (for exmaple, the purple or red region in Figure \ref{fig:eowfig}).


One will see why these states are interpreted as EOW branes by taking the triple-scaling limit in Section \ref{sec:triple}.
We henceforth define 
\begin{align}
    m_{k}^{V,W}
    =\braket{{V}|(\mathsf{T}_{\rm DSSYK}^{\mathcal{N}=0})^k|{W}},
    \label{eq:form2ndssyk2eow}
\end{align}
which we will be interested in from now on.
We depict its chord representation in Figure \ref{fig:eowfig}.

\subsection{$\mathcal{N}=2$ SUSY double-scaled SYK}

A version of the SYK model with $\mathcal{N}=2$ is known in the literature \cite{Sannomiya:2016mnj,Fu:2016vas,Murugan:2017eto,Kanazawa:2017dpd,Garcia-Garcia:2018ruf,Sun:2019yqp}, for which one can also take the double-scaling limit \cite{Berkooz:2020xne,Boruch:2023bte}.
The model consists of $N$ complex fermions satisfying anti-commutation relations
\begin{align}
    \left\{\psi_i,\bar{\psi}_j\right\}=\delta_{i,j},\quad \left\{\psi_i,{\psi}_j\right\}=0,
\end{align}
with Hamiltonian $H_{\rm SUSY}$ given by the anticommutator of two supercharges, $Q$ and $Q^\dagger$
\begin{align}
    H_{\rm SUSY}=\frac{1}{2}\left\{Q,Q^\dagger\right\}.
\end{align}
The supercharges are defined as 
\begin{align}
    \begin{split}
        Q&\equiv i^{{\mathtt{p}}/2} \sum_{1 \le i_1<\cdots <i_{{\mathtt{p}}}\le \mathtt{N} } C_{i_1 i_2 \cdots i_{\mathtt{p}}} \psi_{i_1} \cdots \psi_{i_{\mathtt{p}}},\\ Q^\dagger&\equiv i^{{-\mathtt{p}}/2} \sum_{1 \le i_1<\cdots <i_{{\mathtt{p}}}\le \mathtt{N} } C_{i_1 i_2 \cdots i_{\mathtt{p}}} \bar{\psi}_{i_1} \cdots \bar{\psi}_{i_{\mathtt{p}}},
    \end{split}
\end{align}
where $C_{i_1 i_2 \cdots i_{\mathtt{p}}}$ is, as always, the Gaussian distribution with variance given in \eqref{eq:variance}.

Without any details, let us now review the chord Hilbert space of the double-scaled SYK with $\mathcal{N}=2$ SUSY in the sector with a fixed $R$-charge, $R$.
First of all, even though the basis in chord Hilbert space of the bosonic double-scaled SYK was labelled by the number of chords at each given slice, for the SUSY version each chord is attached a binary choice, the orientation.
Hereafter, let us label the orientation by $\uparrow$ or $\downarrow$, thinking of the time-slice in question as running horizontally.
For example, a basis state with three chords with orientations $\uparrow$, $\uparrow$, $\downarrow$ from left to right, will be denoted as $\ket{\uparrow\uparrow\downarrow,R}$ if it is in the sector with $R$-charge, $R$.

Out of all the orientation choices, only a subset of the configurations are physical -- any other states are deemed null \cite{Berkooz:2020xne,Boruch:2023bte}.
The physical chord Hilbert space $H_{\rm phys}$ is known to be spanned by
\begin{align}
    \left\{\ket{\varnothing,R},\,\ket{(\uparrow\downarrow)^n,R},\,\ket{(\downarrow\uparrow)^n,R},\,\ket{(\uparrow\downarrow)^{n-1}\uparrow,R},\,\ket{(\downarrow\uparrow)^{n-1}\downarrow,R}\right\}.
\end{align}
States with an even number of chords are called bosonic, while odd, fermionic.
The basis are by no means orthonormal, with
\begin{align}
    \begin{gathered}
        \braket{(\uparrow\downarrow)^n,R|(\uparrow\downarrow)^n,R}=\braket{(\downarrow\uparrow)^n,R|(\downarrow\uparrow)^n,R}
        =\texttt{q}^{-n}(\texttt{q}^2;\texttt{q}^2)_{n-1},\\
        \braket{(\uparrow\downarrow)^n,R|(\downarrow\uparrow)^n,R}=-(\texttt{q}^2;\texttt{q}^2)_{n-1}\\
        \braket{(\uparrow\downarrow)^n\uparrow,R|(\uparrow\downarrow)^n\uparrow,R}=\texttt{q}^{-R-n}(\texttt{q}^2;\texttt{q}^2).
    \end{gathered}
\end{align}

In order to write down the transfer matrix $\mathsf{T}_{\rm DSSYK}^{\mathcal{N}=2,R}$, which corresponds to the Hamiltonian, it is useful to introduce a different bosonic basis systems \cite{Aguilar-Gutierrez:2025sqh};
\begin{align}
    \begin{split}
        \ket{b_n}&\equiv \mathtt{q}^n\ket{(\downarrow\uparrow)^n,R}+\ket{(\uparrow\downarrow)^n,R}+\mathtt{q}^R\ket{(\downarrow\uparrow)^{n+1},R}\\
        \ket{b_0}&\equiv \ket{\varnothing}+\mathtt{q}^R\ket{\downarrow\uparrow}\\
        \ket{\bar{b}_n}&\equiv \ket{(\downarrow\uparrow)^n,R}+\mathtt{q}^n\ket{(\uparrow\downarrow)^n,R}+\mathtt{q}^{-R}\ket{(\downarrow\uparrow)^{n+1},R}\\
        \ket{\bar{b}_0}&\equiv \ket{\varnothing}+\mathtt{q}^{-R}\ket{\downarrow\uparrow}
        \end{split},
\end{align}
and the vacuum vector $\ket{\varnothing}$.
Concentrating on the non-BPS sector at $R$-charge, $R$, the transfer matrix becomes
\begin{align}
    \mathsf{T}_{\rm DSSYK}^{\mathcal{N},R}\ket{a_n}
    =
    \sqrt{\frac{1-\mathtt{q}^{2n}}{1-\mathtt{q}^2}}\ket{a_{n-1}}+\frac{\mathtt{q}^{R-\frac{1}{2}}+\mathtt{q}^{-(R-\frac{1}{2})}}{\sqrt{1-\mathtt{q}^2}}\ket{a_n}+\sqrt{\frac{1-\mathtt{q}^{2(n+1)}}{1-\mathtt{q}^2}}\ket{a_{n+1}},
\end{align}
with $\ket{a_n}\equiv \frac{\mathtt{q}^{-n/2}}{\sqrt{(\mathtt{q}^2;\mathtt{q}^2)_n}}\ket{b_n}$.
Equivalently, by setting $q\equiv \mathtt{q}^2$, we have
\begin{align}
    \mathsf{T}_{\rm DSSYK}^{\mathcal{N}=2,R}
    =a^\dagger+a + \frac{q^{\frac{1}{2}(R-\frac{1}{2})}+q^{-\frac{1}{2}(R-\frac{1}{2})}}{\sqrt{1-q}},
\end{align}

The transfer matrix in various $R$-charge sectors can be viewed in a unified way when we introduce another auxiliary Hilbert space, $H_{\rm aux}$.
Let us enlarge our Hilbert space to $H_{\rm chord}\otimes H_{\rm aux}$ and denote the basis state as $\ket{H_n}\otimes \ket{\chi}$, where $\chi\in\mathbb{Z}$.
We then define a new transfer matrix on this enlarged Hilbert space,
\begin{align}
    \mathsf{T}_{\rm DSSYK}\equiv a^\dagger\otimes \mathbbm{1}+a\otimes \mathbbm{1}+\mathbbm{1}\otimes b^\dagger+\mathbbm{1}\otimes b,
    \label{eq:tired}
\end{align}
where the creation/annihilation operators have already been defined in \eqref{eq:aaa} and \eqref{eq:bbb}.
Its relation to the transfer matrix on the chord Hilbert space is quite clear;
we have 
\begin{align}
    \mathsf{T}_{\rm DSSYK}\left(\ket{a_n}\otimes \kket{r}\right)=\left(\mathsf{T}_{\rm DSSYK}^{\mathcal{N}=2,r}\ket{a_n}\right)\otimes \kket{r}, \quad \kket{r}\equiv \sum_{\chi\in\mathbb{Z}}e^{-r\chi}\ket{\chi},
\end{align}
where
\begin{align}
    r\equiv \frac{1}{2}\left(R-\frac{1}{2}\right)
\end{align}
In other words, the supersymmetric as well as the bosonic double-scaled SYK model (which we can get by setting $r=0$ and replacing $q$ with $\mathtt{q}$) can be unified in terms of the new transfer matrix, $\mathsf{T}_{\rm DSSYK}$.
Different $R$-charge sectors of the SUSY double-scaled SYK will be distinguished by distinct momentum sectors inside $H_{\rm aux}$.
Because of this, we will no longer make a precise distinction between SUSY and bosonic double-scaled SYK model.

\subsection{Triple-scaling limit}
\label{sec:triple}

By taking a suitable $q\to 1$ limit of the double-scaled SYK, one recovers one-dimensional Schwarzian dynamics, which can be identified with the low-energy limit of JT gravity.
This limit, called the triple-scaling limit, is again nothing but the limit given in Section \ref{sec:defdef}, \emph{via} the identification that $\lambda\equiv \epsilon\to 0$ \cite{Lin:2022rbf}.
Concretely, our chord basis $\ket{n}\otimes \ket{\chi}$ will be rescaled using $\phi\equiv \lambda n+2\log \lambda$ and $\xi\equiv \lambda \chi$.
Then the low-energy limit is taken by scaling the vacuum-subtracted energy as $\lambda^{3/2}$ --
We can then use \eqref{eq:defdef} and \eqref{eq:defdef111111} to finally obtain the triple-scaling limit of the transfer matrix,
\begin{align}
    \mathsf{T}_{\rm DSSYK}\xrightarrow[\lambda\to 0]{}\frac{4}{\sqrt{\lambda}}-\lambda^{3/2}(\hat{D}_{\rm LQM}+\hat{D}_{\rm free})+O(\lambda^{7/4}),
\end{align}
so that the Hamiltonian for JT gravity on a disk can be written as
\begin{align}
    \hat{D}_{\rm JT}\equiv \hat{D}_{\rm LQM}+\hat{D}_{\rm free}=-\frac{\mathrm{d}^2}{\mathrm{d}\phi^2}+e^{-\phi}-\frac{\mathrm{d}^2}{\mathrm{d}\xi^2}.
\end{align}
The coordinate $\phi$ has an interpretation as the renormalised length of JT gravity on a disk \cite{Lin:2022rbf}, whereas the coordinate $\xi$ is an auxiliary coordinate introduced to keep track of different $R$-charge sectors in $\mathcal{N}=2$ case (or, one can reduce to the bosonic JT gravity when we restrict to the zero momentum sector of $\xi$).
All of this is consistent with the result in \cite{Berkooz:2020xne} (also with \cite{Lin:2022zxd} once we focus on the non-BPS sector).

Let us also discuss what happens to the EOW branes in the $q\to 1$ limit.
By looking at \eqref{eq:coherentlimit}, we simply have to replace\footnote{One can also view the limit at the level of the orthogonality measure as in \cite{Okuyama:2023byh}, proving (4.13) there even when we turn on $\gamma$ and $\delta$.}
\begin{align}
    \begin{split}
        \ket{W}\xrightarrow[\lambda\to 0]{} \cket{u}\equiv \int_{-\infty}^{\infty} \mathrm{d}\phi\, e^{-u\phi}\ket{\phi},\quad \kappa_+(\beta,\delta)\equiv q^u,\\
        \bra{V}\xrightarrow[\lambda\to 0]{} \cbra{v}\equiv \int_{-\infty}^{\infty} \mathrm{d}\phi\, e^{-v\phi}\bra{\phi},\quad \kappa_+(\alpha,\gamma)\equiv q^v.
    \end{split}
    \label{eq:coherentlimitDSSYK}
\end{align}
As before, from now on we will simply write $\ket{{W}}$ to mean $\ket{W}\otimes \sum_{\chi}\ket{\chi}$ and $\cket{u}$ to mean $\cket{u}\otimes \int \mathrm{d}\xi \ket{\xi}$ etc., once we enlarge our Hilbert space.
As discussed in \cite{Okuyama:2023byh}, ${u}$ and ${v}$ correspond to tensions of the EOW brane, $\mu_{{u}}$ and $\mu_{{v}}$ respectively, \emph{via}
\begin{align}
    {u}\equiv \mu_{{u}}+\frac{1}{2}, \quad {v}\equiv \mu_{{v}}+\frac{1}{2}.
\end{align}
As we discussed immediately after \eqref{eq:coherentlimit}, the fact that we require normalisability of $\ket{{V}}$ and $\bra{{W}}$ corresponds to taking ${u}, \, {v}>0$ -- This is nothing but the BF bound for brane tensions in JT gravity \cite{Yang:2018gdb,Kitaev:2018wpr,Okuyama:2023byh}.

Before we close this section, to make contact with the open KPZ stationary measure later, we write the Euclidean evolution operator $e^{-\tau (\hat{D}_{\rm LQM}+\hat{D}_{\rm free})}$ as a scaling limit of double-scaled SYK thermal moments.
Starting from $\braket{W|(\mathsf{T}_{\rm DSSYK})^k|V}$, let us view $\mathsf{T}_{\rm DSSYK}$ as an evolution operator that advances time by one discrete unit.
Let us then take a continuous limit where each time-step is infinitesimal, scaling as $\lambda^2/4$.
To get a finite time propagation from time $\tau$ to $\tau+\delta \tau$, we have
\begin{align}
    (\mathsf{T}_{\rm DSSYK})^{4\delta \tau/\lambda^2}\propto \left(1-\frac{\lambda^2}{4}(\hat{D}_{\rm LQM}+\hat{D}_{\rm free})\right)^{\frac{4}{\lambda^2}\delta \tau}
    \xrightarrow[\lambda\to 0]{}e^{-\delta \tau \hat{D}_{\rm JT}},
    \label{eq:weirdlimit}
\end{align}
recovering the ordinary Euclidean time evolution for JT gravity.

\section{JT/KPZ correspondence}
\label{sec:correspondence}


We now combine the material reviewed in the preceding sections and show that the equivalence between the ASEP stationary measure and the path-integral measure of double-scaled SYK evolved from an EOW brane to another.
We also show that the well known weakly-asymmetric limit for the former and the triple-scaling limit for the latter are the same limit in terms of the correspondence.
The physical consequence of the correspondence is that certain correlators in ASEP or the KPZ equation in the stationary state is exactly the same as those in double-scaled SYK or JT gravity.
In order to avoid complication, we relegate matching the correlators to Appendix \ref{sec:correlators} while we discuss only the path-integral measures here.

\subsection{Double-scaled SYK/ASEP correspondence}

By now it is obvious that there is a correspondence between the double-scaled SYK and ASEP --
\punch{\emph{The transition between two EOW branes parametrised by $(\alpha,\gamma)$ and $(\beta,\delta)$ in double-scaled SYK can be identified with the stationary measure of ASEP with boundary parameters $(\alpha,\gamma)$ and $(\beta,\delta)$.}}
This is in the sense that we have the same discrete path-integral measure for both models:
\begin{align}
    \mathsf{T}_{\rm ASEP}=a^\dagger\otimes \mathbbm{1}+a\otimes \mathbbm{1}+\mathbbm{1}\otimes b^\dagger+\mathbbm{1}\otimes b =\mathsf{T}_{\rm DSSYK}.
\end{align}
We will call the transfer matrix collectively as $\mathsf{T}$.
It acts on $H_{\rm chord}\otimes H_{\rm aux}$, spanned by $\ket{n,\rho}$, where $n,\rho\in\mathbb{Z}$ with $n\geq 0$.
We also start and end with the same initial and final states, $\bra{W}$ and $\ket{V}$.
In this correspondence, it is also clear that the thermal moment $k$ in double-scaled SYK should be identified with the number of sites $N$ in ASEP.\footnote{It is useful to remind ourselves that the correspondence is in between the stationary measure of a Markov process, which is non-unitary viewed as a quantum mechanics, and the path-integral measure of ordinary unitary quantum mechanics. While this seems confusing, the fact that the Markov matrix of ASEP is a non-Hermitian matrix has nothing to do with the fact that its stationary measure is generated by an ordinary Hermitian Hamiltonian. In terms of the Markov process, what we do is to already take the late-time limit, after which we study the random walk that generates the configuration whose time runs in the direction of the originally space-like direction.}


Note that we have taken an unconventional notation in double-scaled SYK (however conventional in ASEP), where the state is represented by a bra instead of a ket, with evolution operators acting from the right rather than the left.
This is only notational as all the operators that we deal with are symmetric operators.

As we have already seen, starting from the same path-integral measure, we get ASEP and double-scaled SYK using a slightly different procedures, respectively;
To get the ASEP stationary measure, we focus on the sum of two processes, $h_{\rm ASEP}=n+\rho$, whereas to get the 
path-integral measure of double-scaled SYK at fixed $R$-charge sector, one simply project to the fixed-momentum sector of $H_{\rm aux}$.

\subsection{JT/KPZ correspondence}

We have also clearly seen that the weakly asymmetric limit of ASEP to KPZ is exactly the same thing as the triple-scaling limit of double-scaled SYK to JT gravity on a disk.
In essence, we can identify the paramter $2/\sqrt{N_{\rm d}}$ in ASEP with $\lambda$ in double-scaled SYK, and the rest follows.
As a result, we claim that \punch{\emph{the transition between two EOW branes with tension $u-\frac{1}{2}$ and $v-\frac{1}{2}$ in JT gravity on a disk can be identified with the stationary measure of open KPZ with Neumann boundary conditions parametrised by $u$ and $v$.}}
As in double-scaled SYK/ASEP correspondence, we have the same Hamiltonian describing both systems, 
\begin{align}
    \hat{D}_{\rm KPZ}=\hat{D}_{\rm LQM}+\hat{D}_{\rm free}=\hat{D}_{\rm JT},
\end{align}
as well as two boundary vectors, $\cket{u}$ and $\cket{v}$, on both sides.
We call the Hamiltonian collectively as $\hat{D}$ from now on.
Here, the Euclidean time between two EOW branes $\beta$ in JT gravity should be identified with the length of the interval $X$ on which the open KPZ is defined.



\section{Conclusions and Outlook}

\label{sec:outlook}


In this paper, we proposed two surprising dualities -- the double-scaled SYK/ASEP and JT/KPZ correspondences.
As these related two seemingly very different topics, let us have an overview of what they were.
In the double-scaled SYK/ASEP correspondence, we had the $N$-th (vacuum-subtracted) thermal moment of the double-scaled SYK on the one hand, and the open ASEP on $N$ sites on the other hand, compared against each other.
We then saw that the open ASEP stationary measure can be thought of as being generated by an {auxiliary spatial transfer matrix}, which turned out to be exactly the same operator as the transfer matrix of the double-scaled SYK in the chord diagram language.
Then the thermal moment was sandwiched by two EOW branes in double-scaled SYK, so that we are interested in the brane-to-brane amplitude.
We also saw that the two parameters characterising each brane corresponds to the boundary inflow and outflow rates at each end of the open ASEP.
Both models had a parameter called $q$; this was related to the number of fermions in the Hamiltonian of the double-scaled SYK, and to the asymmetry hopping rate in the open ASEP.

By taking a suitable $q\to 1$ limit, we found the JT/KPZ correspondence.
The limit to obtain JT gravity from double-scaled SYK is known as the triple-scaling limit, which takes $q\to 1$ while taking the low-energy limit.
The limit to obtain open KPZ equation from the open ASEP, on the other hand, is called the weakly asymmetric limit, which takes $q\to 1$ as well as taking the thermodynamic limit.
They turned out to be exactly the same limit, and this allowed us to argue for the JT/KPZ correspondence.
In the correspondence, the Euclidean evolution for time duration $X$ of JT gravity on a disk was compared against the stationary configuration of the open KPZ equation on an interval of length $X$.
We saw that the stationary measure of the open KPZ can be thought of as a random walk generated by a certain Hamiltonian, which again turned out to be exactly the same as the one describing the dynamics of renormalised wormhole length in JT gravity on a disk.
The Euclidean evolution was sandwiched by two EOW branes with tensions $\mu_u=u-\frac{1}{2}$ and $\mu_{v}=v-\frac{1}{2}$, and these turned out to be related to the parameters appearing in Neumann boundary conditions of the open KPZ height function, $\partial_x h(0)=u$ and $\partial_x h(X)=-v$.

Hoping that these correspondences open a new way of studying quantum gravity, non-equilibrium physics, and relations between them, we list a number of future directions.
First of all, even though we traced back the correspondence to the unifying transfer matrix, $\mathsf{T}=a^\dagger\otimes \mathbbm{1}+a\otimes \mathbbm{1}+\mathbbm{1}\otimes b^\dagger+\mathbbm{1}\otimes b$, it should come from a more fundamental operator, the supercharge, of the $\mathcal{N}=2$ version of the double-scaled SYK.
It would therefore be interesting to trace the correspondence even further back, and interpret the supercharges of SUSY double-scaled SYK in terms of the open ASEP.

Secondly, it would also be interesting to extend the correspondence in various ways.
Exploring the random matrix model viewpoint of the story would be illuminating, as we know that it is deeply related to both models; for example, the Tracy-Widom distribution is known to appear as a probability distribution of the height function at a fixed spatial coordinate \cite{johansson2000shape,prahofer2002scale,sasamoto2010exact,tracy2009asymptotics}, but it also appears as a ground state energy distributions in (low-energy) JT gravity \cite{Saad:2019lba}.
It would be extremely interesting to understand random matrix model, 2D random surfaces (where the other KPZ, the Knizhnik--Polyakov--Zamolodchikov formula \cite{Polyakov:1987zb,Knizhnik:1988ak,DiFrancesco:1993cyw,Distler:1988jt}, appears), and string theory unified under the ASEP and KPZ equation, or {vice versa}.

Relatedly, it would also be interesting to understand if there is any connection to the time-evolution the KPZ equation, not just the stationary measure, to the real-time dynamics of JT gravity.
For example, a crossover from Gaussian to Tracy-Widom fluctuation in the height function is known for ASEP in the weakly asymmetric limit. 
Relating it to, \emph{i.e.}, the spectral form factor would be extremely enlightening.
It would also be interesting to search for a new semi-classical expansion at large number of universes or handles, mimicking the large-charge expansion \cite{Hellerman:2015nra,Monin:2016jmo,Alvarez-Gaume:2016vff,Hellerman:2017efx,Hellerman:2018sjf,Watanabe:2019adh,Gaume:2020bmp,Hellerman:2017veg,Hellerman:2018xpi,Watanabe:2019pdh,Sharon:2020mjs,Watanabe:2022htq,Dodelson:2023uuu,Heckman:2024erd,Watanabe:2025mnc}, which can hopefully be related to various physical quantities on the KPZ side.
Furthermore, it might be interesting to put the correspondence into a broader perspective of the fluid/gravity correspondence \cite{Policastro:2001yc,Janik:2005zt,Bhattacharyya:2007vs,Hubeny:2010ry}.

The duality presented in this paper can be thought of as adding a corner to the triality among double-scaled SYK, Schur half-indices, and 3D $SL(2,\mathbb{C})$ Chern-Simons theory \cite{Gaiotto:2024kze,Lewis:2025qjq,Berkooz:2025ydg}.
For example, one can immediately see that the Schur half-index of $\mathcal{N}=2$ $SU(2)$ gauge theory coupled to $N_{\rm f}=8$ fundamental hypermultiplets can be thought of as the partition function of the open ASEP with $4$ boundary parameters, $(\alpha,\beta,\gamma,\delta)$.
One can also get $N_{\rm f}=6$ by setting $\gamma=0$, $N_{\rm f}=4$ by $\gamma=\delta=0$, $N_{\rm f}=2$ by  $\gamma=\delta=0$ and $\beta =1-q$, $N_{\rm f}=0$ by $\gamma=\delta=0$ and $\alpha=\beta=1-q$.
Furthermore, we can also reproduce the index for the $\mathcal{N}=2^*$ $SU(2)$ SYM, \emph{i.e.}, $\mathcal{N}=2$ $SU(2)$ gauge theory with one adjoint hypermultiplet, when we set $\beta=1-q$ and $\gamma=0$.

As a remark, as the boundary parameters of ASEP will have to become complex to match with Schur half-indices, we might as well consider the XXZ model on an interval with non-diagonal boundary conditions, which can be obtained as a similarity transform of ASEP \cite{de2005bethe}.
Although the boundary rates being positive is a physical requirement for ASEP, such a requirement is no longer necessary when we consider the XXZ model with possibly non-unitary boundary conditions.
Therefore it would be interesting to more extensively look at the duality between Schur half-indices and 3D $SL(2,\mathbb{C})$ Chern-Simons theory from the viewpoint of the integrable XXZ spin chain.
It would also be possible to generalise the correspondence to general $SU(n)$ gauge theories, in which case we might obtain a variation of ASEP with $n-1$ species \cite{crampe2016matrix}.

Finally, it would be extremely fruitful to understand the relation to de Sitter space, of which there are several versions, in the language of ASEP \cite{Susskind:2021esx,Susskind:2022dfz,Susskind:2022bia,Susskind:2023hnj,Rahman:2023pgt,Rahman:2024iiu,Sekino:2025bsc,Narovlansky:2023lfz,Verlinde:2024znh,Verlinde:2024zrh,Tietto:2025oxn,Blommaert:2024whf,Blommaert:2024ymv,Okuyama:2025hsd,Aguilar-Gutierrez:2025hty,Narovlansky:2025tpb}.
It would, for example, be possible to identify fake temperatures in the semi-classical limit of double-scaled SYK relevant for such a discussion \cite{Lin:2023trc}; it might be worthwhile to use another representation for the DEHP algebra given in \cite{enaud2004large}, as this could make such a $q\to 1$ limit more uniform.

\section*{Acknowledgements}
The author thanks Micha Berkooz, Trivko Kukolj, Kazumi Okuyama, Josef Seitz, and Masahito Yamazaki for valuable discussions.
This work is supported by a Grant-in-Aid for JSPS Fellows No.~22KJ1777, a Grant-in-Aid for Early-Career Scientists No.~25K17387, and by a MEXT KAKENHI Grant No.~24H00957.

\appendix

\section{Some $q$-combinatorics}
\label{sec:q-pol}


We collect some notations on $q$-combinatorics to be used to the main text.
The $q$-Pochhammer symbols are defined as
\begin{align}
    (a;q)_n\equiv \prod_{i=1}^{n}(1-aq^{i-1})\\
    (q;q)_n\equiv \prod_{i=1}^{n}(1-q^i).
\end{align}
We will also use a shorthand notation for producs of $q$-Pochhammer symbols, such as
\begin{align}
    (a,b;q)_n&\equiv (a;q)_n(b;q)_n\\
    (a,e^{\pm 2i\theta};q)&\equiv (a;q)_n(e^{2i\theta};q)_n(e^{-2i\theta};q)_n
\end{align}
The $q$-binomial is defined as
\begin{align}
    \qbinom{L}{N}\equiv \frac{(q;q)_L}{(q;q)_N(q;q)_{L-N}}
\end{align}
The continuous $q$-Hermite polynomial is defined as
\begin{align}
    H_n(\cos\theta|q)\equiv \sum_{k=0}^{n}\qbinom{L}{N}e^{i(n-2k)\theta},
\end{align}
and they satisfy a recursion relation,
\begin{align}
    2x H_n(x|q)=H_{n+1}(x|q)+(1-q^2)H_{n-1}(x|q)
    \label{eq:rec}
\end{align}
We also introduce an identity for $H_n(\cos\theta|q)$ to be used in the main text,
\begin{align}
    \sum_{n=0}^{\infty}H_n(\cos\theta|q)H_n(\cos\phi|q)\frac{t^n}{(q;q)_n}=\frac{(t^2;q)_\infty}{(te^{i(\pm\theta\pm \phi)};q)_\infty}.
    \label{eq:qdefcom2}
\end{align}

\section{Matching correlators}
\label{sec:correlators}


\subsection{Double-scaled SYK/ASEP correspondence}

We start from the stationary distribution of ASEP, \eqref{eq:probint}.
It is customary to study the Laplace-transformed version, which can be computed, using the DEHP algebra, as
\begin{align}
    \Braket{\prod_{i=1}^N {t_i}^{\uptau_i}}\equiv \sum_{\vec{\uptau}}P(\vec{\uptau}){t_1}^{\uptau_1}{t_2}^{\uptau_2}\cdots {t_N}^{\uptau_L}=\frac{\Pi(t_1,t_2,\dots, t_N)}{\Pi(1,1,\dots, 1)},
\end{align}
where 
\begin{align}
    \begin{split}
        \Pi(t_1,t_2,\dots, t_N)&\equiv \braket{W|\prod_{j=1}^{N}\left(\EE +t_j\DD \right)|V}\\
        &=\braket{W|\prod_{j=1}^{N}\left[(t_j)^{-\hat{N}/2}\left(\frac{1+t_j}{\sqrt{1-q}}+\sqrt{t_j}(a^\dagger +a)\right)(t_j)^{\hat{N}/2}\right]|V}.
    \end{split}
    \label{eq:qqqqq}
\end{align}
The second equality used the fact that tri-diagonal matrices can be symmetrised, \emph{i.e.},
\begin{align}
    t^{\hat{N}/2}(\EE + t \DD)t^{-\hat{N}/2} =\frac{1+t}{\sqrt{1-q}}+\sqrt{t}(a^\dagger + a).
\end{align}
The reason why inserting $t_i^{\uptau_i}$ changes the ``vacuum energy'' of the evolution operator can easily be understood in terms of the unified transfer matrix, $\mathsf{T}\equiv a^\dagger\otimes \mathbbm{1}+a\otimes \mathbbm{1}+\mathbbm{1}\otimes b^\dagger+\mathbbm{1}\otimes b$;
it is simply because \emph{via} $h_{\rm ASEP}=n+\chi$, its exponential gives momentum to the auxiliary Hilbert space, $H_{\rm aux}$.
(See the main text for definitions.)

We can now write down the $(n+1)$-point function on the stationary measure of the ASEP height function.
We have
\begin{align}
    \begin{split}
        &{\Braket{e^{-\sum_{j=1}^{n+1}s_j\left(h_{\rm ASEP}(k_j)-h_{\rm ASEP}(k_{j-1})\right)}}}\\
        &\qquad\qquad =e^{\sum_{j=1}^{n+1}s_j(k_j-k_{j-1})}\frac{\Pi(\overbrace{e^{-2s_1},\dots,e^{-2s_1}}^{k_1-k_0},\dots,\overbrace{e^{-2s_{n+1}},\dots, e^{-2s_{n+1}}}^{k_{n+1}-k_{n}})}{\Pi(1,\dots,1)}
    \end{split}
    \label{eq:mpltaseop}
\end{align}
where we have set $k_0\equiv 0$ and $k_{n+1}\equiv N$.
The exponential in front comes from the definition of $h_{\rm ASEP}$ in relation to $\uptau$.
We also define $c_j\equiv s_j-s_{j+1}$, where $s_{n+2}\equiv 0$. Notably, we have $s_1\equiv c_1+\cdots c_n$ and $s_{n+1}=c_{n+1}$.
It is also immediate to see that
\begin{align}
    \label{eq:VWASEP}
    \begin{split}
            &{\Theta}(\overbrace{e^{-2s_1},\dots,e^{-2s_1}}^{k_1-k_0},\dots,\overbrace{e^{-2s_{n+1}},\dots, e^{-2s_{n+1}}}^{k_{n+1}-k_{{n}}})\\
            &\qquad \qquad
        \equiv 
        e^{\sum_{j=1}^{{n+1}}s_j(k_j-k_{j-1})}\Pi(\overbrace{e^{-2s_1},\dots,e^{-2s_1}}^{k_1-k_0},\dots,\overbrace{e^{-2s_{n+1}},\dots, e^{-2s_{n+1}}}^{k_{n+1}-k_{n}})\\
    & \qquad \qquad =
    \braket{W|
    e^{s_1\hat{N}}
    \prod_{j=1}^{{n+1}}\left[
    \left(\frac{2\cosh(s_j)}{\sqrt{1-q}}+a^\dagger +a\right)^{k_j-k_{j-1}}e^{-c_j\hat{N}}
    \right]
    |V}.
    \end{split}
\end{align}

We can furthermore rewrite this as follows, with redefinitions $i_j\equiv k_{j+1}-k_j$,
\begin{align}
    \label{eq:VWASEP2}
    \begin{split}
            &{\Theta}(\overbrace{e^{-2s_1},\dots,e^{-2s_1}}^{i_0},\dots,\overbrace{e^{-2s_{n+1}},\dots, e^{-2s_{n+1}}}^{i_{n}})\\
    = &
    \braket{W|
    e^{s_1(\hat{N}+\hat{\chi})}
    \mathsf{T}^{i_0}e^{-c_1(\hat{N}+\hat{\chi})}\cdots e^{-c_n(\hat{N}+\hat{\chi})}
    \mathsf{T}^{i_{n}}e^{-s_{n+1}(\hat{N}+\hat{\chi})}
    |V},
    \end{split}
\end{align}
by using \eqref{eq:tired} and $\hat{\chi}\ket{\chi}={\chi}\ket{\chi}$.
This is just rewriting everything in terms of $H_{\rm chord}\otimes H_{\rm aux}$, so that the shift in vacuum-energy in the evolution operator can be absorbed into the change in momentum in the auxiliary Hilbert space.

Let us now discuss operator insertions in SUSY double-scaled SYK.
We consider insertions of random operators of the form,
\begin{align}
    M_A\equiv i^{{\mathtt{p}}_A/2} \sum _{1 \le i_1<\cdots <i_{{\mathtt{p}}_A}\le N } J_{i_1 i_2 \cdots i_{{\mathtt{p}}_A}} \psi_{i_1} \cdots \psi_{i_{{\mathtt{p}}_A}},
    \label{eq:randomop}
\end{align}
where $J_{i_1 i_2 \cdots i_{{\mathtt{p}}_A}}$ is again Gaussian random with variance
\begin{align}
    \left\langle J_{i_1 i_2 \cdots i_{{\mathtt{p}}_A}}J_{j_1 j_2 \cdots j_{{\mathtt{p}}_A}}\right\rangle={\binom{\mathtt{N}}{{{\mathtt{p}}_A}}}^{-1}\delta_{i_1,j_1}\cdots \delta_{i_{{\mathtt{p}}_A},j_{{\mathtt{p}}_A}}.
\end{align}
It is known that the parameters admit a natural double-scaling limit,
\begin{align}
    \lambda_A\equiv \frac{2{\mathtt{p}}{\cdot{\mathtt{p}}_A}}{\mathtt{N}}=\fixed, \quad 
    {q}_A\equiv e^{-\lambda_A},
    \quad \lambda_A\equiv \ell_A\lambda.
\end{align}
In this limit, a contraction of indices in $J_{i_1 i_2 \cdots i_{p_A}}$ corresponds to chord that represents matter, which contributes as $q_A$ when crossed with an ordinary chord.
Different matter chords can cross too, but we are not interested in such cases in this paper -- The correlators without any matter chord crossings will be called \emph{uncrossed} correlators hereafter.

We are hereafter interested in the uncrossed $2n$-point function of the following form,
\begin{align}
    m_{i_0i_1\cdots i_{n}}\equiv \biggl\langle\braket{\tilde{V}|\mathsf{H}^{i_1}
    \contraction[3ex]{}{\bar{M}_{1}}{\mathsf{H}^{i_1}\bar{M}_{2}\mathsf{H}^{i_{2}}\bar{M}_3\cdots M_3M_2}{M_1}
    \contraction[2ex]{\bar{M}_{1}\mathsf{H}^{i_1}}{\bar{M}_{2}}{\mathsf{H}^{i_{2}}\bar{M}_3\cdots M_3}{M_2}
    \contraction[1ex]{\bar{M}_{1}\mathsf{H}^{i_1}\bar{M}_{2}\mathsf{H}^{i_{2}}}{\bar{M}_3}{\cdots }{M_3}
    \bar{M}_{1}\mathsf{H}^{i_2}\bar{M}_{2}\mathsf{H}^{i_{3}}\bar{M}_3\cdots M_3M_2M_1
    |\tilde{W}}
    \biggr\rangle,
    \label{eq:oper}
\end{align}
where the contraction symbol $\contraction{}{\bar{M_k}}{}{{M_k}}\bar{M_{k}}M_{k}$ means the matter chord connecting two random operators.
We denote the length of the random operators $M_k$ as $p_k$, while the double-scaling parameters are denoted as $\lambda_k$, $q_k$, and $\ell_k$.
Note further that the operator $M_k$ has an $R$-charge of $\ell_k$, in our normalisation.
The states $\ket{\tilde{W}}$ and $\bra{\tilde{V}}$ are the EOW branes, whose definition in terms of the chord Hilbert space is given in the main body of the text.

\begin{figure}[t]
    \centering
    \begin{tikzpicture}[scale=1]
        \draw (244:2cm) arc (244:116:2cm);
        \draw (64:2cm) arc (64:-64:2cm);
        \draw [ultra thick, BlueViolet] (64:2cm) arc (-26:-154:1cm);
        \draw [ultra thick, BrickRed] (-64:2cm) arc (26:154:1cm);
        \draw [BlueViolet] (40:2cm) arc (-40:-140:2cm);
        \filldraw[BlueViolet] (40:2cm) circle (3pt);
        \filldraw[BlueViolet] (140:2cm) circle (3pt);
        \draw [BrickRed] (-40:2cm) arc (40:140:2cm);
        \filldraw[BrickRed] (-40:2cm) circle (3pt);
        \filldraw[BrickRed] (-140:2cm) circle (3pt);
        \draw [YellowGreen] (0:2cm) -- (180:2cm);
        \filldraw[YellowGreen] (0:2cm) circle (3pt);
        \filldraw[YellowGreen] (180:2cm) circle (3pt);
        \node at (90:2.1cm) {\rotatebox{-90}{\large $\langle \tilde{V} \!\mid$}};
        \node at (-90:2.1cm) {\rotatebox{-90}{\large $\mid\! \tilde{W} \rangle$}};
        \node[above right] at (40:2cm) {$\bar{M}_3$};
        \node[above left] at (140:2cm) {$M_3$};
        \node[right] at (0:2cm) {$\bar{M}_2$};
        \node[left] at (180:2cm) {$M_2$};
        \node[below left] at (-140:2cm) {$M_1$};
        \node[below right] at (-40:2cm) {$\bar{M}_1$};
    \end{tikzpicture}
    \caption{A chord representation of an uncrossed six-point function of the form \eqref{eq:oper}.
    }
    \label{fig:opening2}
\end{figure}
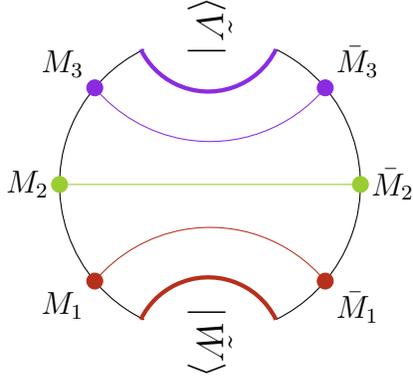

We depict the chord diagram representation of a six-point uncrossed correlator in Figure \ref{fig:opening2}.
Note that we have used the fact that the bi-local operator $\contraction{}{\bar{M}_k}{}{M_k}\bar{M}_{k}M_{k}$ commutes with $\mathsf{H}$, which was used to shift them all the way to the right.
This is because it makes no difference to close or open a chord on the left or the right arc in the doubled Hilbert space picture.

As discussed in \cite{},
the bilocal operator $\contraction{}{\bar{M}_k}{}{M_k}\bar{M}_{k}M_{k}$
acts on the chord Hilbert space as $(q_k)^{\hat{N}_{\rm chord}}$, where $\hat{N}_{\rm chord}$ counts the number of chords.
In SUSY double-scaled SYK, such an operator acts on basis states $\ket{a_n,r}\equiv \ket{a_n}\otimes \kket{r}$ as
\begin{align}
    (q_k)^{\hat{N}_{\rm chord}}\ket{a_n,r}=(q_k)^n\ket{a_n,r+\ell_k},
\end{align}
so that $\hat{N}_{\rm chord}=\hat{N}\otimes \hat{\chi}$, where $\hat{N}\ket{a_n}\equiv n\ket{a_n}$ and $\hat{\chi}\ket{\chi}=\chi\ket{\chi}$.
This immediately gives us an expression for the uncrossed $2n$-point function in the double-scaled SYK,
\begin{align}
    m_{i_1\cdots i_n}
    =\braket{\tilde{V}|(\mathsf{T}_{\rm ASEP})^{i_1}q^{\ell_1(\hat{N}+\hat{\rho})}(\mathsf{T}_{\rm ASEP})^{i_2}\cdots q^{\ell_n(\hat{N}+\hat{\rho})}(\mathsf{T}_{\rm ASEP})^{i_{n}}|\tilde{W}}.
    \label{eq:form2ndssyk}
\end{align}

Now it is clear that $\Theta(t_1,\dots, t_N)$ can be exactly matched with $m_{i_1\cdots i_{n}}$, by comparing \eqref{eq:form2ndssyk2eow} and \eqref{eq:qqqqq}.
As one can immediately see, insertions in both models can be written as $q^{\ell(\hat{N}+\hat{\chi})}$; $e^{-c_j}$ in ASEP simply corresponds to $q^{\ell_j}$ in double-scaled SYK.
The only difference between the two is the two operators just before and after the coherent states in ASEP.
However, as we have argued in \eqref{eq:trickman}, we have a relation $e^{-c_{n+1}(\hat{N}+\hat{\chi})}\ket{W}\otimes \kket{r}=\ket{\tilde{W}}\otimes \kket{r+c_{n+1}}$, for a suitable choice of another coherent state $\ket{\tilde{W}}$,
and hence will not cause problems in matching the correlators.

\subsection{JT/KPZ correspondence}

By taking a weakly asymmetric limit of the open ASEP stationary measure, one can obtain the stationary measure for open KPZ.
Again, we are interested in the multi-point Laplace transform of the stationary measure for the height function, which is expressed as ${\braket{e^{-\sum_{j=1}^{n}\sigma_j\left(h(x_j)-h(x_{j-1})\right)}}}$,
with $x_0\equiv 0$ and $x_n\equiv T$.
As is the case with ASEP, we define $\kappa_j\equiv \sigma_j-\sigma_{j+1}$, where $\sigma_{n+1}=0$.
Notably we have $\sigma_1\equiv \kappa_1+\dots + \kappa_n$ and $\sigma_n\equiv \kappa_n$.

Let us compute the above Laplace transform by taking the weakly asymmetric limit of open ASEP.
The weakly asymmetric limit \eqref{eq:waseplimit} suggests that we have
\begin{align}
    \lim_{N\to \infty}
    {\Braket{e^{-\sum_{j=1}^{n}\frac{\sigma_j}{\sqrt{N_{\mathrm{d}}}}\left(h_{\mathrm{ASEP}}(N_{\mathrm{d}}x_j)-h_{\mathrm{ASEP}}(N_{\mathrm{d}}x_{j-1})\right)}}}
    ={\Braket{e^{-\sum_{j=1}^{n}\sigma_j\left(h(x_j)-h(x_{j-1})\right)}}}.
\end{align}
Therefore by using \eqref{eq:mpltaseop}, we have that
\begin{align}
    \begin{split}
        {\Braket{e^{-\sum_{j=1}^{n}\sigma_j\left(h(x_j)-h(x_{j-1})\right)}}}
    =\lim_{N\to\infty}\frac{\Theta(\vec{\sigma})}{\Theta(\vec{0})},
    \end{split}
\end{align}
where we have defined
\begin{align}
    \Theta(\vec{\sigma})\equiv {\Theta}(\overbrace{e^{-2\sigma_1/\sqrt{N_{\rm d}}},\dots,e^{-2\sigma_1/\sqrt{N_{\rm d}}}}^{{N_{\rm d}x_1}-{N_{\rm d}x_{0}}},\dots,\overbrace{e^{-2\sigma_n/\sqrt{N_{\rm d}}},\dots, e^{-2\sigma_n/\sqrt{N_{\rm d}}}}^{{N_{\rm d}x_n}-{N_{\rm d}x_{n-1}}}).
\end{align}

We now compute $\Theta(\vec{\sigma})$ by using \eqref{eq:VWASEP} and
plugging \eqref{eq:defdef} and \eqref{eq:coherentlimit} into \eqref{eq:VWASEP}, we conclude that
\begin{align}
    \tilde{\Theta}(\vec{\sigma})\xrightarrow[N_{\rm d}\to \infty]{}
    e^{\frac{1}{4}\sum_{j=1}^{n}{\sigma_j}^2(x_j-x_{j-1})}
    \cbraket{u|e^{\sigma_1\hat{\phi}}
    \prod_{j=1}^{n}\left[e^{-(x_j-x_{j-1})\hat{D}_{\rm LQM}}e^{-{\kappa_j}\hat{\phi}}\right]
    |v}
    ,
\end{align}
where we have defined $\tilde{\Theta}(\vec{\sigma})\equiv \left(\frac{4}{\sqrt{1-q}}\right)^{-\beta N_{\rm d}}\Theta(\vec{\sigma})$,
stripping off a diverging prefactor.
Note also that, by definition, we can rewrite the above by using $\cbra{u}e^{\sigma_1\hat{\phi}}=\cbra{u-\sigma_1}$ and $e^{-\kappa_n\hat{\phi}}\cket{v}=\cket{v+\kappa_n}$.

To sum up, we are able to express the $(n+1)$-point Laplace transform of the open KPZ stationary measure in terms of the $n$-point function of the Liouville quantum mechanics.
Concretely, we have found that
\begin{align}
    \begin{split}
        &{\Braket{e^{-\sum_{j=1}^{n+1}\sigma_j\left(h(x_j)-h(x_{j-1})\right)}}}
        =
        e^{\frac{1}{4}\sum_{j=1}^{n+1}{\sigma_j}^2(x_j-x_{j-1})}\\
        &\quad
        \times 
        \frac{\cbraket{u-\sigma_1|e^{-x_1 D_{\rm LQM}}e^{-\kappa_1 \hat{\phi}}e^{-(x_2-x_1) D_{\rm LQM}}
        \cdots e^{-\kappa_n \hat{\phi}}e^{-(T -x_{n}) D_{\rm LQM}}
        |v+\sigma_n}
        }{\cbraket{u|e^{-\beta D_{\rm LQM}}|v}}.
    \end{split}
\end{align}
One can also write it as 
\begin{align}
    \begin{split}
        &{\Braket{e^{-\sum_{j=1}^{n+1}\sigma_j\left(h(x_j)-h(x_{j-1})\right)}}}\\
        = &
        \frac{\cbraket{u-\sigma_1|e^{-x_1 (D_{\rm LQM}+D_{\rm free})}e^{-\kappa_1 (\hat{\phi}+\hat{\xi})}\cdots e^{-\kappa_n (\hat{\phi}+\hat{\xi})}e^{-(X-x_{n}) (D_{\rm LQM}+D_{\rm free})}
        |v+\sigma_n}
        }{\cbraket{u|e^{-X (D_{\rm LQM}+D_{\rm free})}|v}},
    \end{split}
\end{align}

By now, it is clear that this is nothing but the $2n$-point function of (super-)JT gravity, evaluated between two EOW branes.
Indeed by looking at \eqref{eq:form2ndssyk2eow} and taking the triple-scaling limit, we see that it can be written as
\begin{align}
    \begin{split}
        &\braket{v|\mathcal{O}_1(\tau_1)\cdots \mathcal{O}_n(\tau_n)|u}_\beta\\
        =&
        \frac{\cbraket{u|e^{-\tau_1 (D_{\rm LQM}+D_{\rm free})}e^{-\kappa_1 (\hat{\phi}+\hat{\xi})}\cdots e^{-\kappa_n (\hat{\phi}+\hat{\xi})}e^{-(\beta-\tau_{n}) (D_{\rm LQM}+D_{\rm free})}
        |v}
        }{\cbraket{u|e^{-\beta (D_{\rm LQM}+D_{\rm free})}|v}},
    \end{split}
\end{align}
where $\mathcal{O}_k(\tau_k)$ corresponds to inserting a bilocal operator $\contraction{}{\bar{M}_k}{}{M_k}\bar{M}_{k}M_{k}$ at Euclidean time $\tau_k$.

\section{Askey--Wilson transfer matrix}
\label{sec:fin}


There are other representations of the DEHP algebra one can use to compute $\braket{\prod_{i=1}^N {t_i}^{\uptau_i}}$.
Of particular interest to us is the Uchiyama--Sasamoto--Wadachi (USW) representation \cite{uchiyama2004asymmetric}, where we use the one given explicitly in (3.8) of \cite{wang2024askey} where $A$, $B$, $C$, and $D$ given there are exactly the ones given in this paper as \eqref{eq:ASEPparameters}.
It is the generalization of the operator defined in (5.7) of \cite{Berkooz:2025ydg}. 
The advantage of this representation is that the boundary vectors $\bra{V}$ and $\ket{W}$ are simply written as $(1,0,\dots)$ and $(1,0,\dots)^T$, and so the analysis is not just limited to the ASEP fan region.
The operators in the USW representations will be given a subscript USW from now on.

As discussed in \cite{Berkooz:2025ydg}, the transfer matrix $\DD_{\rm USW}+\EE_{\rm USW}$ can be thought of as the Hamiltonian of the double-scaled SYK in the ``dual channel'', which views amplitudes between the two EOW branes as vacuum-to-vacuum amplitudes of systems with boundary conditions on the two ends.\footnote{After taking the triple-scaling limit, $(\DD+\EE)_{\rm USW}$ becomes the Schr\"odinger operator with P\"oschl--Teller potential, which may be related to \cite{Griguolo:2026eoi}.}
By using it, we can write the unnormalised Laplace transform of the ASEP stationary measure in the USW representation,
\begin{align}
    \Pi(t_1,\dots,t_N)\equiv \braket{0|\prod_{i=1}^N(\DD+t_i\EE)_{\rm USW}|0},
\end{align}
where $t_i\equiv e^{-2s_i}$.

What we would point out is that chord basis representation and the USW representation is related in a very simple, algebraic way, without proof.
First of all, the vacuum state can be written as
\begin{align}
    \bra{0}=\int \mathrm{d}\theta\, \braket{V|(t_1)^{\hat{N}/2}|\theta_1}\braket{\theta_1|(t_1)^{-\hat{N}/2}|W}\bra{\theta_1^{[t_1]}},
\end{align}
where $\bra{\theta_1^{[t]}}$ is an eigenstate of the operator $(\DD+t\EE)_{\rm USW}$,
\begin{align}
    \bra{\theta_1^{[t]}}(\DD+t\EE)_{\rm USW}=(1+t+2\sqrt{t}\cos\theta)\bra{\theta_1^{[t]}},
\end{align}
while $\ket{\theta}$ is the eigenstate of the operator $\DD+\EE$ as usual.
Furthermore, we have
\begin{align}
    \braket{\theta_1^{[t_1]}|\theta_2^{[t_2]}}=\frac{\braket{\theta_1|(t_2)^{-\hat{N}/2}|W}}{\braket{\theta_1|(t_1)^{-\hat{N}/2}|W}}\times \braket{\theta_1|(t_1/t_2)^{-\hat{N}/2}|\theta_2}.
\end{align}
These are enough to relate the USW representation of $\Pi(\vec{t})$ to its chord basis representation by cancelling $\braket{\theta_1|(t_i)^{-\hat{N}/2}|W}$ and replacing it with $\braket{\theta_1|(t_{i+1})^{-\hat{N}/2}|W}$ at each step of $(\DD+t_i\EE)_{\rm USW}$.

\bibliographystyle{JHEP}
\bibliography{refs-new,hep}

\providecommand{\href}[2]{#2}\begingroup\raggedright\begin{thebibliography}{100}

\bibitem{maunuksela1997kinetic}
J.~Maunuksela, M.~Myllys, O.-P.~K{\"a}hk{\"o}nen, J.~Timonen, N.~Provatas, M.~Alava et~al., \emph{Kinetic roughening in slow combustion of paper}, {\emph{Physical review letters} {\bfseries 79} (1997) 1515}.

\bibitem{kardar1986dynamic}
M.~Kardar, G.~Parisi and Y.-C.~Zhang, \emph{Dynamic scaling of growing interfaces}, {\emph{Physical Review Letters} {\bfseries 56} (1986) 889}.

\bibitem{wakita1997self}
J.-i.~Wakita, H.~Itoh, T.~Matsuyama and M.~Matsushita, \emph{Self-affinity for the growing interface of bacterial colonies}, {\emph{Journal of the Physical Society of Japan} {\bfseries 66} (1997) 67}.

\bibitem{hallatschek2007genetic}
O.~Hallatschek, P.~Hersen, S.~Ramanathan and D.R.~Nelson, \emph{Genetic drift at expanding frontiers promotes gene segregation}, {\emph{Proceedings of the National Academy of Sciences} {\bfseries 104} (2007) 19926}.

\bibitem{bertini1997stochastic}
L.~Bertini and G.~Giacomin, \emph{Stochastic burgers and kpz equations from particle systems}, {\emph{Communications in mathematical physics} {\bfseries 183} (1997) 571}.

\bibitem{funaki2015kpz}
T.~Funaki and J.~Quastel, \emph{Kpz equation, its renormalization and invariant measures}, {\emph{Stochastic Partial Differential Equations: Analysis and Computations} {\bfseries 3} (2015) 159}.

\bibitem{sasamoto2010one}
T.~Sasamoto and H.~Spohn, \emph{One-dimensional kardar-parisi-zhang equation: an exact solution and its universality}, {\emph{Physical review letters} {\bfseries 104} (2010) 230602}.

\bibitem{amir2011probability}
G.~Amir, I.~Corwin and J.~Quastel, \emph{Probability distribution of the free energy of the continuum directed random polymer in 1+ 1 dimensions}, {\emph{Communications on pure and applied mathematics} {\bfseries 64} (2011) 466}.

\bibitem{calabrese2011exact}
P.~Calabrese and P.~Le~Doussal, \emph{Exact solution for the kardar-parisi-zhang equation with flat initial conditions}, {\emph{Physical review letters} {\bfseries 106} (2011) 250603}.

\bibitem{imamura2012exact}
T.~Imamura and T.~Sasamoto, \emph{Exact solution for the stationary kardar-parisi-zhang equation}, {\emph{Physical review letters} {\bfseries 108} (2012) 190603}.

\bibitem{imamura2013stationary}
T.~Imamura and T.~Sasamoto, \emph{Stationary correlations for the 1d kpz equation}, {\emph{Journal of Statistical Physics} {\bfseries 150} (2013) 908}.

\bibitem{corwin2018open}
I.~Corwin and H.~Shen, \emph{Open asep in the weakly asymmetric regime}, {\emph{Communications on Pure and Applied Mathematics} {\bfseries 71} (2018) 2065}.

\bibitem{parekh2019kpz}
S.~Parekh, \emph{The kpz limit of asep with boundary}, {\emph{Communications in Mathematical Physics} {\bfseries 365} (2019) 569}.

\bibitem{yang2025kpz}
K.~Yang, \emph{Kpz equation from open asep with general boundary asymmetry}, {\emph{arXiv preprint arXiv:2507.11537} (2025) }.

\bibitem{derrida1998exactly}
B.~Derrida, \emph{An exactly soluble non-equilibrium system: the asymmetric simple exclusion process}, {\emph{Physics Reports} {\bfseries 301} (1998) 65}.

\bibitem{macdonald1968kinetics}
C.T.~MacDonald, J.H.~Gibbs and A.C.~Pipkin, \emph{Kinetics of biopolymerization on nucleic acid templates}, {\emph{Biopolymers: Original Research on Biomolecules} {\bfseries 6} (1968) 1}.

\bibitem{schutz2001exactly}
G.M.~Sch{\"u}tz, \emph{Exactly solvable models for many-body systems far from equilibrium},  in \emph{Phase transitions and critical phenomena}, vol.~19, pp.~1--251, Elsevier (2001).

\bibitem{schadschneider2010stochastic}
A.~Schadschneider, D.~Chowdhury and K.~Nishinari, \emph{Stochastic transport in complex systems: from molecules to vehicles}, Elsevier (2010).

\bibitem{derrida1993exact}
B.~Derrida, M.R.~Evans, V.~Hakim and V.~Pasquier, \emph{Exact solution of a 1d asymmetric exclusion model using a matrix formulation}, {\emph{Journal of Physics A: Mathematical and General} {\bfseries 26} (1993) 1493}.

\bibitem{uchiyama2004asymmetric}
M.~Uchiyama, T.~Sasamoto and M.~Wadati, \emph{Asymmetric simple exclusion process with open boundaries and askey--wilson polynomials}, {\emph{Journal of Physics A: Mathematical and General} {\bfseries 37} (2004) 4985}.

\bibitem{sasamoto1999one}
T.~Sasamoto, \emph{One-dimensional partially asymmetric simple exclusion process with open boundaries: orthogonal polynomials approach}, {\emph{Journal of Physics A: Mathematical and General} {\bfseries 32} (1999) 7109}.

\bibitem{Jackiw:1984je}
R.~Jackiw, \emph{{Lower Dimensional Gravity}}, \href{https://doi.org/10.1016/0550-3213(85)90448-1}{\emph{Nucl. Phys. B} {\bfseries 252} (1985) 343}.

\bibitem{Teitelboim:1983ux}
C.~Teitelboim, \emph{{Gravitation and Hamiltonian Structure in Two Space-Time Dimensions}}, \href{https://doi.org/10.1016/0370-2693(83)90012-6}{\emph{Phys. Lett. B} {\bfseries 126} (1983) 41}.

\bibitem{Maldacena:2016upp}
J.~Maldacena, D.~Stanford and Z.~Yang, \emph{{Conformal symmetry and its breaking in two dimensional Nearly Anti-de-Sitter space}}, \href{https://doi.org/10.1093/ptep/ptw124}{\emph{PTEP} {\bfseries 2016} (2016) 12C104} [\href{https://arxiv.org/abs/1606.01857}{{\ttfamily 1606.01857}}].

\bibitem{Bagrets:2016cdf}
D.~Bagrets, A.~Altland and A.~Kamenev, \emph{{Sachdev{\textendash}Ye{\textendash}Kitaev model as Liouville quantum mechanics}}, \href{https://doi.org/10.1016/j.nuclphysb.2016.08.002}{\emph{Nucl. Phys. B} {\bfseries 911} (2016) 191} [\href{https://arxiv.org/abs/1607.00694}{{\ttfamily 1607.00694}}].

\bibitem{Lam:2018pvp}
H.T.~Lam, T.G.~Mertens, G.J.~Turiaci and H.~Verlinde, \emph{{Shockwave S-matrix from Schwarzian Quantum Mechanics}}, \href{https://doi.org/10.1007/JHEP11(2018)182}{\emph{JHEP} {\bfseries 11} (2018) 182} [\href{https://arxiv.org/abs/1804.09834}{{\ttfamily 1804.09834}}].

\bibitem{Goel:2018ubv}
A.~Goel, H.T.~Lam, G.J.~Turiaci and H.~Verlinde, \emph{{Expanding the Black Hole Interior: Partially Entangled Thermal States in SYK}}, \href{https://doi.org/10.1007/JHEP02(2019)156}{\emph{JHEP} {\bfseries 02} (2019) 156} [\href{https://arxiv.org/abs/1807.03916}{{\ttfamily 1807.03916}}].

\bibitem{Berkooz:2022fso}
M.~Berkooz, N.~Brukner, S.F.~Ross and M.~Watanabe, \emph{{Going beyond ER=EPR in the SYK model}}, \href{https://doi.org/10.1007/JHEP08(2022)051}{\emph{JHEP} {\bfseries 08} (2022) 051} [\href{https://arxiv.org/abs/2202.11381}{{\ttfamily 2202.11381}}].

\bibitem{Lin:2022rbf}
H.W.~Lin, \emph{{The bulk Hilbert space of double scaled SYK}}, \href{https://doi.org/10.1007/JHEP11(2022)060}{\emph{JHEP} {\bfseries 11} (2022) 060} [\href{https://arxiv.org/abs/2208.07032}{{\ttfamily 2208.07032}}].

\bibitem{Sachdev:1992fk}
S.~Sachdev and J.~Ye, \emph{{Gapless spin fluid ground state in a random, quantum Heisenberg magnet}}, \href{https://doi.org/10.1103/PhysRevLett.70.3339}{\emph{Phys. Rev. Lett.} {\bfseries 70} (1993) 3339} [\href{https://arxiv.org/abs/cond-mat/9212030}{{\ttfamily cond-mat/9212030}}].

\bibitem{Maldacena:2016hyu}
J.~Maldacena and D.~Stanford, \emph{{Remarks on the Sachdev-Ye-Kitaev model}}, \href{https://doi.org/10.1103/PhysRevD.94.106002}{\emph{Phys. Rev. D} {\bfseries 94} (2016) 106002} [\href{https://arxiv.org/abs/1604.07818}{{\ttfamily 1604.07818}}].

\bibitem{Kitaev2015Talks}
A.~Kitaev, ``A simple model of quantum holography.'' KITP talks, Apr 7 and May 27, 2015.

\bibitem{Berkooz:2018qkz}
M.~Berkooz, P.~Narayan and J.~Simon, \emph{{Chord diagrams, exact correlators in spin glasses and black hole bulk reconstruction}}, \href{https://doi.org/10.1007/JHEP08(2018)192}{\emph{JHEP} {\bfseries 08} (2018) 192} [\href{https://arxiv.org/abs/1806.04380}{{\ttfamily 1806.04380}}].

\bibitem{Berkooz:2018jqr}
M.~Berkooz, M.~Isachenkov, V.~Narovlansky and G.~Torrents, \emph{{Towards a full solution of the large N double-scaled SYK model}}, \href{https://doi.org/10.1007/JHEP03(2019)079}{\emph{JHEP} {\bfseries 03} (2019) 079} [\href{https://arxiv.org/abs/1811.02584}{{\ttfamily 1811.02584}}].

\bibitem{Berkooz:2024lgq}
M.~Berkooz and O.~Mamroud, \emph{{A cordial introduction to double scaled SYK}}, \href{https://doi.org/10.1088/1361-6633/ada889}{\emph{Rept. Prog. Phys.} {\bfseries 88} (2025) 036001} [\href{https://arxiv.org/abs/2407.09396}{{\ttfamily 2407.09396}}].

\bibitem{Okuyama:2023byh}
K.~Okuyama, \emph{{End of the world brane in double scaled SYK}}, \href{https://doi.org/10.1007/JHEP08(2023)053}{\emph{JHEP} {\bfseries 08} (2023) 053} [\href{https://arxiv.org/abs/2305.12674}{{\ttfamily 2305.12674}}].

\bibitem{Watanabe:2024vad}
M.~Watanabe, \emph{{On perturbation around closed exclusion processes}}, \href{https://doi.org/10.21468/SciPostPhys.17.3.092}{\emph{SciPost Phys.} {\bfseries 17} (2024) 092} [\href{https://arxiv.org/abs/2406.02675}{{\ttfamily 2406.02675}}].

\bibitem{Berkooz:2025ydg}
M.~Berkooz, T.~Kukolj and J.~Seitz, \emph{{Comments on Class S(YK)}},  \href{https://arxiv.org/abs/2507.12524}{{\ttfamily 2507.12524}}.

\bibitem{Okuyama:2025ebn}
K.~Okuyama, \emph{{Holographic tensor network for double-scaled SYK}}, \href{https://doi.org/10.21468/SciPostPhys.19.4.083}{\emph{SciPost Phys.} {\bfseries 19} (2025) 083} [\href{https://arxiv.org/abs/2503.23003}{{\ttfamily 2503.23003}}].

\bibitem{Berkooz:2020xne}
M.~Berkooz, N.~Brukner, V.~Narovlansky and A.~Raz, \emph{{The double scaled limit of Super--Symmetric SYK models}}, \href{https://doi.org/10.1007/JHEP12(2020)110}{\emph{JHEP} {\bfseries 12} (2020) 110} [\href{https://arxiv.org/abs/2003.04405}{{\ttfamily 2003.04405}}].

\bibitem{barraquand2021steady}
G.~Barraquand and P.L.~Doussal, \emph{Steady state of the kpz equation on an interval and liouville quantum mechanics}, {\emph{arXiv preprint arXiv:2105.15178} (2021) }.

\bibitem{corwin2024stationary}
I.~Corwin and A.~Knizel, \emph{Stationary measure for the open kpz equation}, {\emph{Communications on Pure and Applied Mathematics} {\bfseries 77} (2024) 2183}.

\bibitem{barraquand2023stationary}
G.~Barraquand and P.~Le~Doussal, \emph{Stationary measures of the kpz equation on an interval from enaud--derrida’s matrix product ansatz representation}, {\emph{Journal of Physics A: Mathematical and Theoretical} {\bfseries 56} (2023) 144003}.

\bibitem{Gaiotto:2024kze}
D.~Gaiotto and H.~Verlinde, \emph{{SYK-Schur duality: double scaled SYK correlators from $ \mathcal{N} $ = 2 supersymmetric gauge theory}}, \href{https://doi.org/10.1007/JHEP06(2025)163}{\emph{JHEP} {\bfseries 06} (2025) 163} [\href{https://arxiv.org/abs/2409.11551}{{\ttfamily 2409.11551}}].

\bibitem{Lewis:2025qjq}
O.~Lewis, M.~Mezei, M.~Sacchi and S.~Schafer-Nameki, \emph{{Schur Connections: Chord Counting, Line Operators, and Indices}},  \href{https://arxiv.org/abs/2506.17384}{{\ttfamily 2506.17384}}.

\bibitem{Gorsky:2018jdj}
A.S.~Gorsky, S.K.~Nechaev and A.F.~Valov, \emph{{On statistical models on super trees}}, \href{https://doi.org/10.1007/JHEP08(2018)123}{\emph{JHEP} {\bfseries 08} (2018) 123} [\href{https://arxiv.org/abs/1801.03067}{{\ttfamily 1801.03067}}].

\bibitem{Gorsky:2020lma}
A.~Gorsky, S.~Nechaev and A.~Valov, \emph{{Lifshitz tails at spectral edge and holography with a finite cutoff}}, \href{https://doi.org/10.1007/JHEP04(2021)080}{\emph{JHEP} {\bfseries 04} (2021) 080} [\href{https://arxiv.org/abs/2011.04600}{{\ttfamily 2011.04600}}].

\bibitem{Gorsky:2024xnl}
A.~Gorsky, S.~Nechaev and A.~Valov, \emph{{KPZ scaling from the Krylov space}}, \href{https://doi.org/10.1007/JHEP09(2024)021}{\emph{JHEP} {\bfseries 09} (2024) 021} [\href{https://arxiv.org/abs/2406.02782}{{\ttfamily 2406.02782}}].

\bibitem{Derrida:1992vu}
B.~Derrida, M.R.~Evans, V.~Hakim and V.~Pasquier, \emph{{Exact solution of a 1d asymmetric exclusion model using a matrix formulation}}, {\emph{J. Phys. A} {\bfseries 26} (1993) 1493}.

\bibitem{Arik:1973vg}
M.~Arik and D.D.~Coon, \emph{{Hilbert Spaces of Analytic Functions and Generalized Coherent States}}, \href{https://doi.org/10.1063/1.522937}{\emph{J. Math. Phys.} {\bfseries 17} (1976) 524}.

\bibitem{bryc2017asymmetric}
W.~Bryc and J.~Weso{\l}owski, \emph{Asymmetric simple exclusion process with open boundaries and quadratic harnesses}, {\emph{Journal of Statistical Physics} {\bfseries 167} (2017) 383}.

\bibitem{wang2024askey}
Y.~Wang, J.~Weso{\l}owski and Z.~Yang, \emph{Askey--wilson signed measures and open asep in the shock region}, {\emph{International Mathematics Research Notices} {\bfseries 2024} (2024) 11104}.

\bibitem{Sannomiya:2016mnj}
N.~Sannomiya, H.~Katsura and Y.~Nakayama, \emph{{Supersymmetry breaking and Nambu-Goldstone fermions with cubic dispersion}}, \href{https://doi.org/10.1103/PhysRevD.95.065001}{\emph{Phys. Rev. D} {\bfseries 95} (2017) 065001} [\href{https://arxiv.org/abs/1612.02285}{{\ttfamily 1612.02285}}].

\bibitem{Fu:2016vas}
W.~Fu, D.~Gaiotto, J.~Maldacena and S.~Sachdev, \emph{{Supersymmetric Sachdev-Ye-Kitaev models}}, \href{https://doi.org/10.1103/PhysRevD.95.026009}{\emph{Phys. Rev. D} {\bfseries 95} (2017) 026009} [\href{https://arxiv.org/abs/1610.08917}{{\ttfamily 1610.08917}}].

\bibitem{Murugan:2017eto}
J.~Murugan, D.~Stanford and E.~Witten, \emph{{More on Supersymmetric and 2d Analogs of the SYK Model}}, \href{https://doi.org/10.1007/JHEP08(2017)146}{\emph{JHEP} {\bfseries 08} (2017) 146} [\href{https://arxiv.org/abs/1706.05362}{{\ttfamily 1706.05362}}].

\bibitem{Kanazawa:2017dpd}
T.~Kanazawa and T.~Wettig, \emph{{Complete random matrix classification of SYK models with $\mathcal{N}=0$, $1$ and $2$ supersymmetry}}, \href{https://doi.org/10.1007/JHEP09(2017)050}{\emph{JHEP} {\bfseries 09} (2017) 050} [\href{https://arxiv.org/abs/1706.03044}{{\ttfamily 1706.03044}}].

\bibitem{Garcia-Garcia:2018ruf}
A.M.~Garc{\'\i}a-Garc{\'\i}a, Y.~Jia and J.J.M.~Verbaarschot, \emph{{Universality and Thouless energy in the supersymmetric Sachdev-Ye-Kitaev Model}}, \href{https://doi.org/10.1103/PhysRevD.97.106003}{\emph{Phys. Rev. D} {\bfseries 97} (2018) 106003} [\href{https://arxiv.org/abs/1801.01071}{{\ttfamily 1801.01071}}].

\bibitem{Sun:2019yqp}
F.~Sun and J.~Ye, \emph{{Periodic Table of the Ordinary and Supersymmetric Sachdev-Ye-Kitaev Models}}, \href{https://doi.org/10.1103/PhysRevLett.124.244101}{\emph{Phys. Rev. Lett.} {\bfseries 124} (2020) 244101} [\href{https://arxiv.org/abs/1905.07694}{{\ttfamily 1905.07694}}].

\bibitem{Boruch:2023bte}
J.~Boruch, H.W.~Lin and C.~Yan, \emph{{Exploring supersymmetric wormholes in $ \mathcal{N} $ = 2 SYK with chords}}, \href{https://doi.org/10.1007/JHEP12(2023)151}{\emph{JHEP} {\bfseries 12} (2023) 151} [\href{https://arxiv.org/abs/2308.16283}{{\ttfamily 2308.16283}}].

\bibitem{Aguilar-Gutierrez:2025sqh}
S.E.~Aguilar-Gutierrez, \emph{{Evolution With(out) Time: Relational Holography {\&} BPS Complexity Growth in $\mathcal{N}=2$ Double-Scaled SYK}},  \href{https://arxiv.org/abs/2510.11777}{{\ttfamily 2510.11777}}.

\bibitem{Lin:2022zxd}
H.W.~Lin, J.~Maldacena, L.~Rozenberg and J.~Shan, \emph{{Looking at supersymmetric black holes for a very long time}}, \href{https://doi.org/10.21468/SciPostPhys.14.5.128}{\emph{SciPost Phys.} {\bfseries 14} (2023) 128} [\href{https://arxiv.org/abs/2207.00408}{{\ttfamily 2207.00408}}].

\bibitem{Yang:2018gdb}
Z.~Yang, \emph{{The Quantum Gravity Dynamics of Near Extremal Black Holes}}, \href{https://doi.org/10.1007/JHEP05(2019)205}{\emph{JHEP} {\bfseries 05} (2019) 205} [\href{https://arxiv.org/abs/1809.08647}{{\ttfamily 1809.08647}}].

\bibitem{Kitaev:2018wpr}
A.~Kitaev and S.J.~Suh, \emph{{Statistical mechanics of a two-dimensional black hole}}, \href{https://doi.org/10.1007/JHEP05(2019)198}{\emph{JHEP} {\bfseries 05} (2019) 198} [\href{https://arxiv.org/abs/1808.07032}{{\ttfamily 1808.07032}}].

\bibitem{johansson2000shape}
K.~Johansson, \emph{Shape fluctuations and random matrices}, {\emph{Communications in mathematical physics} {\bfseries 209} (2000) 437}.

\bibitem{prahofer2002scale}
M.~Pr{\"a}hofer and H.~Spohn, \emph{Scale invariance of the png droplet and the airy process}, {\emph{Journal of statistical physics} {\bfseries 108} (2002) 1071}.

\bibitem{sasamoto2010exact}
T.~Sasamoto and H.~Spohn, \emph{Exact height distributions for the kpz equation with narrow wedge initial condition}, {\emph{Nuclear Physics B} {\bfseries 834} (2010) 523}.

\bibitem{tracy2009asymptotics}
C.A.~Tracy and H.~Widom, \emph{Asymptotics in asep with step initial condition}, {\emph{Communications in Mathematical Physics} {\bfseries 290} (2009) 129}.

\bibitem{Saad:2019lba}
P.~Saad, S.H.~Shenker and D.~Stanford, \emph{{JT gravity as a matrix integral}},  \href{https://arxiv.org/abs/1903.11115}{{\ttfamily 1903.11115}}.

\bibitem{Polyakov:1987zb}
A.M.~Polyakov, \emph{{Quantum Gravity in Two-Dimensions}}, \href{https://doi.org/10.1142/S0217732387001130}{\emph{Mod. Phys. Lett. A} {\bfseries 2} (1987) 893}.

\bibitem{Knizhnik:1988ak}
V.G.~Knizhnik, A.M.~Polyakov and A.B.~Zamolodchikov, \emph{{Fractal Structure of 2D Quantum Gravity}}, \href{https://doi.org/10.1142/S0217732388000982}{\emph{Mod. Phys. Lett. A} {\bfseries 3} (1988) 819}.

\bibitem{DiFrancesco:1993cyw}
P.~Di~Francesco, P.H.~Ginsparg and J.~Zinn-Justin, \emph{{2-D Gravity and random matrices}}, \href{https://doi.org/10.1016/0370-1573(94)00084-G}{\emph{Phys. Rept.} {\bfseries 254} (1995) 1} [\href{https://arxiv.org/abs/hep-th/9306153}{{\ttfamily hep-th/9306153}}].

\bibitem{Distler:1988jt}
J.~Distler and H.~Kawai, \emph{{Conformal Field Theory and 2D Quantum Gravity}}, \href{https://doi.org/10.1016/0550-3213(89)90354-4}{\emph{Nucl. Phys. B} {\bfseries 321} (1989) 509}.

\bibitem{Hellerman:2015nra}
S.~Hellerman, D.~Orlando, S.~Reffert and M.~Watanabe, \emph{{On the CFT Operator Spectrum at Large Global Charge}}, \href{https://doi.org/10.1007/JHEP12(2015)071}{\emph{JHEP} {\bfseries 12} (2015) 071} [\href{https://arxiv.org/abs/1505.01537}{{\ttfamily 1505.01537}}].

\bibitem{Monin:2016jmo}
A.~Monin, D.~Pirtskhalava, R.~Rattazzi and F.K.~Seibold, \emph{{Semiclassics, Goldstone Bosons and CFT data}}, \href{https://doi.org/10.1007/JHEP06(2017)011}{\emph{JHEP} {\bfseries 06} (2017) 011} [\href{https://arxiv.org/abs/1611.02912}{{\ttfamily 1611.02912}}].

\bibitem{Alvarez-Gaume:2016vff}
L.~Alvarez-Gaume, O.~Loukas, D.~Orlando and S.~Reffert, \emph{{Compensating strong coupling with large charge}}, \href{https://doi.org/10.1007/JHEP04(2017)059}{\emph{JHEP} {\bfseries 04} (2017) 059} [\href{https://arxiv.org/abs/1610.04495}{{\ttfamily 1610.04495}}].

\bibitem{Hellerman:2017efx}
S.~Hellerman, N.~Kobayashi, S.~Maeda and M.~Watanabe, \emph{{A Note on Inhomogeneous Ground States at Large Global Charge}}, \href{https://doi.org/10.1007/JHEP10(2019)038}{\emph{JHEP} {\bfseries 10} (2019) 038} [\href{https://arxiv.org/abs/1705.05825}{{\ttfamily 1705.05825}}].

\bibitem{Hellerman:2018sjf}
S.~Hellerman, N.~Kobayashi, S.~Maeda and M.~Watanabe, \emph{{Observables in inhomogeneous ground states at large global charge}}, \href{https://doi.org/10.1007/JHEP08(2021)079}{\emph{JHEP} {\bfseries 08} (2021) 079} [\href{https://arxiv.org/abs/1804.06495}{{\ttfamily 1804.06495}}].

\bibitem{Watanabe:2019adh}
M.~Watanabe, \emph{{Chern-Simons-matter theories at large baryon number}}, \href{https://doi.org/10.1007/JHEP10(2021)245}{\emph{JHEP} {\bfseries 10} (2021) 245} [\href{https://arxiv.org/abs/1904.09815}{{\ttfamily 1904.09815}}].

\bibitem{Gaume:2020bmp}
L.{\'A}.~Gaum{\'e}, D.~Orlando and S.~Reffert, \emph{{Selected topics in the large quantum number expansion}}, \href{https://doi.org/10.1016/j.physrep.2021.08.001}{\emph{Phys. Rept.} {\bfseries 933} (2021) 1} [\href{https://arxiv.org/abs/2008.03308}{{\ttfamily 2008.03308}}].

\bibitem{Hellerman:2017veg}
S.~Hellerman, S.~Maeda and M.~Watanabe, \emph{{Operator Dimensions from Moduli}}, \href{https://doi.org/10.1007/JHEP10(2017)089}{\emph{JHEP} {\bfseries 10} (2017) 089} [\href{https://arxiv.org/abs/1706.05743}{{\ttfamily 1706.05743}}].

\bibitem{Hellerman:2018xpi}
S.~Hellerman, S.~Maeda, D.~Orlando, S.~Reffert and M.~Watanabe, \emph{{Universal correlation functions in rank 1 SCFTs}}, \href{https://doi.org/10.1007/JHEP12(2019)047}{\emph{JHEP} {\bfseries 12} (2019) 047} [\href{https://arxiv.org/abs/1804.01535}{{\ttfamily 1804.01535}}].

\bibitem{Watanabe:2019pdh}
M.~Watanabe, \emph{{Accessing large global charge via the $\epsilon$-expansion}}, \href{https://doi.org/10.1007/JHEP04(2021)264}{\emph{JHEP} {\bfseries 04} (2021) 264} [\href{https://arxiv.org/abs/1909.01337}{{\ttfamily 1909.01337}}].

\bibitem{Sharon:2020mjs}
A.~Sharon and M.~Watanabe, \emph{{Transition of Large $R$-Charge Operators on a Conformal Manifold}}, \href{https://doi.org/10.1007/JHEP01(2021)068}{\emph{JHEP} {\bfseries 01} (2021) 068} [\href{https://arxiv.org/abs/2008.01106}{{\ttfamily 2008.01106}}].

\bibitem{Watanabe:2022htq}
M.~Watanabe, \emph{{Stability analysis of a non-unitary CFT}}, \href{https://doi.org/10.1007/JHEP11(2023)042}{\emph{JHEP} {\bfseries 11} (2023) 042} [\href{https://arxiv.org/abs/2203.08843}{{\ttfamily 2203.08843}}].

\bibitem{Dodelson:2023uuu}
M.~Dodelson, S.~Hellerman, M.~Watanabe and M.~Yamazaki, \emph{{Integrability of large-charge sectors in generic 2D EFTs}}, \href{https://doi.org/10.1007/JHEP08(2024)166}{\emph{JHEP} {\bfseries 08} (2024) 166} [\href{https://arxiv.org/abs/2310.01823}{{\ttfamily 2310.01823}}].

\bibitem{Heckman:2024erd}
J.J.~Heckman, A.~Sharon and M.~Watanabe, \emph{{6D large charge and 2D Virasoro blocks}}, \href{https://doi.org/10.1103/PhysRevD.111.045002}{\emph{Phys. Rev. D} {\bfseries 111} (2025) 045002} [\href{https://arxiv.org/abs/2409.05944}{{\ttfamily 2409.05944}}].

\bibitem{Watanabe:2025mnc}
M.~Watanabe, \emph{{Large-charge R{\'e}nyi entropy}},  \href{https://arxiv.org/abs/2506.10072}{{\ttfamily 2506.10072}}.

\bibitem{Policastro:2001yc}
G.~Policastro, D.T.~Son and A.O.~Starinets, \emph{{The Shear viscosity of strongly coupled N=4 supersymmetric Yang-Mills plasma}}, \href{https://doi.org/10.1103/PhysRevLett.87.081601}{\emph{Phys. Rev. Lett.} {\bfseries 87} (2001) 081601} [\href{https://arxiv.org/abs/hep-th/0104066}{{\ttfamily hep-th/0104066}}].

\bibitem{Janik:2005zt}
R.A.~Janik and R.B.~Peschanski, \emph{{Asymptotic perfect fluid dynamics as a consequence of Ads/CFT}}, \href{https://doi.org/10.1103/PhysRevD.73.045013}{\emph{Phys. Rev. D} {\bfseries 73} (2006) 045013} [\href{https://arxiv.org/abs/hep-th/0512162}{{\ttfamily hep-th/0512162}}].

\bibitem{Bhattacharyya:2007vs}
S.~Bhattacharyya, S.~Lahiri, R.~Loganayagam and S.~Minwalla, \emph{{Large rotating AdS black holes from fluid mechanics}}, \href{https://doi.org/10.1088/1126-6708/2008/09/054}{\emph{JHEP} {\bfseries 09} (2008) 054} [\href{https://arxiv.org/abs/0708.1770}{{\ttfamily 0708.1770}}].

\bibitem{Hubeny:2010ry}
V.E.~Hubeny and M.~Rangamani, \emph{{A Holographic view on physics out of equilibrium}}, \href{https://doi.org/10.1155/2010/297916}{\emph{Adv. High Energy Phys.} {\bfseries 2010} (2010) 297916} [\href{https://arxiv.org/abs/1006.3675}{{\ttfamily 1006.3675}}].

\bibitem{de2005bethe}
J.~De~Gier and F.H.~Essler, \emph{Bethe ansatz solution of the asymmetric exclusion process with open boundaries}, {\emph{Physical review letters} {\bfseries 95} (2005) 240601}.

\bibitem{crampe2016matrix}
N.~Cramp{\'e}, M.~Evans, K.~Mallick, E.~Ragoucy and M.~Vanicat, \emph{Matrix product solution to a 2-species tasep with open integrable boundaries}, {\emph{Journal of Physics A: Mathematical and Theoretical} {\bfseries 49} (2016) 475001}.

\bibitem{Susskind:2021esx}
L.~Susskind, \emph{{Entanglement and Chaos in De Sitter Space Holography: An SYK Example}}, \href{https://doi.org/10.22128/jhap.2021.455.1005}{\emph{JHAP} {\bfseries 1} (2021) 1} [\href{https://arxiv.org/abs/2109.14104}{{\ttfamily 2109.14104}}].

\bibitem{Susskind:2022dfz}
L.~Susskind, \emph{{Scrambling in Double-Scaled SYK and De Sitter Space}},  \href{https://arxiv.org/abs/2205.00315}{{\ttfamily 2205.00315}}.

\bibitem{Susskind:2022bia}
L.~Susskind, \emph{{De Sitter Space, Double-Scaled SYK, and the Separation of Scales in the Semiclassical Limit}}, \href{https://doi.org/10.22128/jhap.2024.920.1103}{\emph{JHAP} {\bfseries 5} (2025) 1} [\href{https://arxiv.org/abs/2209.09999}{{\ttfamily 2209.09999}}].

\bibitem{Susskind:2023hnj}
L.~Susskind, \emph{{De Sitter Space has no Chords. Almost Everything is Confined.}}, \href{https://doi.org/10.22128/jhap.2023.661.1043}{\emph{JHAP} {\bfseries 3} (2023) 1} [\href{https://arxiv.org/abs/2303.00792}{{\ttfamily 2303.00792}}].

\bibitem{Rahman:2023pgt}
A.A.~Rahman and L.~Susskind, \emph{{Comments on a Paper by Narovlansky and Verlinde}},  \href{https://arxiv.org/abs/2312.04097}{{\ttfamily 2312.04097}}.

\bibitem{Rahman:2024iiu}
A.A.~Rahman and L.~Susskind, \emph{{$p$-Chords, Wee-Chords, and de Sitter Space}},  \href{https://arxiv.org/abs/2407.12988}{{\ttfamily 2407.12988}}.

\bibitem{Sekino:2025bsc}
Y.~Sekino and L.~Susskind, \emph{{Double-Scaled SYK, QCD, and the Flat Space Limit of de Sitter Space}},  \href{https://arxiv.org/abs/2501.09423}{{\ttfamily 2501.09423}}.

\bibitem{Narovlansky:2023lfz}
V.~Narovlansky and H.~Verlinde, \emph{{Double-scaled SYK and de Sitter holography}}, \href{https://doi.org/10.1007/JHEP05(2025)032}{\emph{JHEP} {\bfseries 05} (2025) 032} [\href{https://arxiv.org/abs/2310.16994}{{\ttfamily 2310.16994}}].

\bibitem{Verlinde:2024znh}
H.~Verlinde, \emph{{Double-scaled SYK, chords and de Sitter gravity}}, \href{https://doi.org/10.1007/JHEP03(2025)076}{\emph{JHEP} {\bfseries 03} (2025) 076} [\href{https://arxiv.org/abs/2402.00635}{{\ttfamily 2402.00635}}].

\bibitem{Verlinde:2024zrh}
H.~Verlinde and M.~Zhang, \emph{{SYK correlators from 2D Liouville-de Sitter gravity}}, \href{https://doi.org/10.1007/JHEP05(2025)053}{\emph{JHEP} {\bfseries 05} (2025) 053} [\href{https://arxiv.org/abs/2402.02584}{{\ttfamily 2402.02584}}].

\bibitem{Tietto:2025oxn}
D.~Tietto and H.~Verlinde, \emph{{A microscopic model of de Sitter spacetime with an observer}},  \href{https://arxiv.org/abs/2502.03869}{{\ttfamily 2502.03869}}.

\bibitem{Blommaert:2024whf}
A.~Blommaert, A.~Levine, T.G.~Mertens, J.~Papalini and K.~Parmentier, \emph{{An entropic puzzle in periodic dilaton gravity and DSSYK}}, \href{https://doi.org/10.1007/JHEP07(2025)093}{\emph{JHEP} {\bfseries 07} (2025) 093} [\href{https://arxiv.org/abs/2411.16922}{{\ttfamily 2411.16922}}].

\bibitem{Blommaert:2024ymv}
A.~Blommaert, T.G.~Mertens and J.~Papalini, \emph{{The dilaton gravity hologram of double-scaled SYK}}, \href{https://doi.org/10.1007/JHEP06(2025)050}{\emph{JHEP} {\bfseries 06} (2025) 050} [\href{https://arxiv.org/abs/2404.03535}{{\ttfamily 2404.03535}}].

\bibitem{Okuyama:2025hsd}
K.~Okuyama, \emph{{de Sitter JT gravity from double-scaled SYK}}, \href{https://doi.org/10.1007/JHEP08(2025)181}{\emph{JHEP} {\bfseries 08} (2025) 181} [\href{https://arxiv.org/abs/2505.08116}{{\ttfamily 2505.08116}}].

\bibitem{Aguilar-Gutierrez:2025hty}
S.E.~Aguilar-Gutierrez, \emph{{Symmetry sectors in chord space and relational holography in the DSSYK. Lessons from branes, wormholes, and de Sitter space}}, \href{https://doi.org/10.1007/JHEP10(2025)044}{\emph{JHEP} {\bfseries 10} (2025) 044} [\href{https://arxiv.org/abs/2506.21447}{{\ttfamily 2506.21447}}].

\bibitem{Narovlansky:2025tpb}
V.~Narovlansky, \emph{{Towards a microscopic description of de Sitter dynamics}},  \href{https://arxiv.org/abs/2506.02109}{{\ttfamily 2506.02109}}.

\bibitem{Lin:2023trc}
H.W.~Lin and D.~Stanford, \emph{{A symmetry algebra in double-scaled SYK}}, \href{https://doi.org/10.21468/SciPostPhys.15.6.234}{\emph{SciPost Phys.} {\bfseries 15} (2023) 234} [\href{https://arxiv.org/abs/2307.15725}{{\ttfamily 2307.15725}}].

\bibitem{enaud2004large}
C.~Enaud and B.~Derrida, \emph{Large deviation functional of the weakly asymmetric exclusion process}, {\emph{Journal of statistical physics} {\bfseries 114} (2004) 537}.

\bibitem{Griguolo:2026eoi}
L.~Griguolo, J.~Papalini, L.~Russo, D.~Seminara and A.~Tarana, \emph{{Quantum JT Gravity in a box as a P{\"o}schl-Teller Scattering Problem}},  \href{https://arxiv.org/abs/2607.01385}{{\ttfamily 2607.01385}}.

\end{thebibliography}\endgroup

\end{document}